\documentclass[12pt,preprint]{aastex}
\newcommand{\kms}{\,km~s$^{-1}$}

\newcommand{\Msun}{\mbox{\,$M_{\odot}$}}

\newcommand{\Sersic}{S\'ersic}
\def\spose#1{\hbox to 0pt{#1\hss}}
\def\simlt{\mathrel{\spose{\lower 3pt\hbox{$\mathchar"218$}}
     \raise 2.0pt\hbox{$\mathchar"13C$}}}
\def\simgt{\mathrel{\spose{\lower 3pt\hbox{$\mathchar"218$}}
     \raise 2.0pt\hbox{$\mathchar"13E$}}}

\newcommand{\ngal}{101}
\newcommand{\ngalnoHI}{88}
\newcommand{\ngaldiff}{13}

\newcommand{\fgas}{0.6}
\newcommand{\maxfgas}{95}

\begin{document}




\title{The Baryon Content of Extremely Low Mass Dwarf Galaxies}


\author{M.\ Geha\altaffilmark{1}}
\affil{The Observatories of the Carnegie Institute of Washington,
        813 Santa Barbara St., Pasadena, CA 91101. mgeha@ociw.edu}
\altaffiltext{1}{Hubble Fellow.}

\author{M.\ R.\ Blanton, M.\ Masjedi} \affil{Center for Cosmology and
Particle Physics, New York University, New York, NY
10003. blanton@nyu.edu, morad.masjedi@physics.nyu.edu}

\author{A.\ A.\ West} 
\affil{Astronomy Department, University of
California, 601 Campbell Hall, Berkeley, CA 94720. awest@astron.berkeley.edu}


\begin{abstract}
\renewcommand{\thefootnote}{\fnsymbol{footnote}}

We investigate the gas content and baryonic Tully-Fisher relationship
for extremely low luminosity dwarf galaxies in the absolute magnitude
range $-13.5>M_r-5\log_{10} h_{70} >-16$.  The sample is selected from
the Sloan Digital Sky Survey and consists of \ngal\ galaxies for which
we have obtained follow-up HI observations using the Arecibo
Observatory and Green Bank Telescope.  This represents the largest
homogeneous sample of dwarf galaxies at low luminosities with
well-measured HI and optical properties.  The sample spans a range of
environments, from dense groups to truly isolated galaxies. The
average neutral gas fraction is $\langle f_{\rm gas}\rangle=$\,\fgas,
significantly exceeding that of typical gas-rich galaxies at higher
luminosities.  Dwarf galaxies are therefore less efficient at turning
gas into stars over their lifetimes.  The strong environmental
dependence of the gas fraction distribution demonstrates that while
internal processes can reduce the gas fractions to roughly $f_{\rm
  gas} = 0.4$, external processes are required to fully remove gas
from a dwarf galaxy.  The average rotational velocity of our sample is
$\langle v_{\rm rot}\rangle=\langle$W$20_{i,t}/2\rangle=50$ \kms\
based on HI line-widths.  In this luminosity range, the optical
Tully-Fisher relationship has significantly more scatter compared to
the baryonic relationship.  Including more massive galaxies from the
literature, we fit a baryonic Tully-Fisher slope of $M_{\rm baryon}
\propto v_{\rm rot}^{3.70\pm 0.15}$.  This slope compares well with
CDM models that assume an equal baryon to dark matter ratio at all
masses.  While gas stripping or other processes may modify the baryon
to dark matter ratio for dwarfs in the densest environments, the
majority of dwarf galaxies in our sample have not preferentially lost
significant baryonic mass relative to more massive galaxies.

\end{abstract}




\section{Introduction}\label{sec_intro}
\renewcommand{\thefootnote}{\fnsymbol{footnote}}

The properties of low mass dwarf galaxies, and their differences from
luminous galaxies, provide important clues to understanding both
cosmology and galaxy formation.  As probes of galaxy formation, dwarf
galaxies should be extremely sensitive to processes such as star
formation feedback, ram pressure stripping, and tidal stripping.
These processes are often invoked to explain properties of relatively
luminous galaxies and should even more dramatically affect dwarf
galaxies, which have smaller masses and shallower potential wells.  As
probes of cosmology, the number of dwarf galaxies has provided a
challenge for cosmological models.  

The current Cold Dark Matter model with a cosmological constant
($\Lambda$CDM; \citealt{spergel06a}) makes robust predictions for the
number of dark matter halos as a function of mass \citep{sheth01a}.
However, extending this model to predict the number density of
observed galaxies requires including physical effects such as gas
cooling, star-formation, supernova feedback, etc., that are too
complex to follow exactly in numerical predictions.  It is clear that
the number of galaxies as a function of luminosity is much shallower
than the prediction for the number as a function of halo mass
\citep{benson03a}. The most extreme example of this problem is the
``substructure problem'' in the Local Group
\citep{klypin99a,moore99a}.  If we assume galaxies are associated
one-to-one with dark matter halos and that $\Lambda$CDM is correct,
the mass-to-light ratios of galaxies therefore must decline as a function of
host halo mass.  That is, the efficiency of forming stars and/or
retaining gas must be much smaller for low mass galaxies.

A number of physical models have been proposed to explain high
mass-to-light ratios in low mass galaxies, including photoevaporation
of gas during reionization \citep{shapiro04a,mayer04a}, suppression of
gas accretion during reionization
\citep{gnedin00a,bullock00a,benson03a}, feedback from supernovae
\citep{dekel03a,nagashima04a}, and ram pressure stripping of gas
\citep{mori00a}.  Some of these processes only occur for dwarf
galaxies near a luminous galaxy and not for isolated ones
\citep{kravtsov04a}.  The relative importance of these processes in
the evolution of dwarf galaxies has significant implications for
galaxies of all sizes at both low and high redshift.  For example, the
importance of galactic winds in low mass galaxies will strongly affect
our understanding of the mass-metallicity relation
\citep{tremonti04a,lee06a} and the enrichment of the intergalactic
medium at high redshift \citep{scannapieco05a}.  By comparing the gas
content and Tully-Fisher relationships of dwarfs as a function of
environment, it should be possible to determine the importance of
``external'' environmental effects like stripping relative to
``internal'' effects such as feedback.

Historically, extremely low luminosity nearby galaxies have been
difficult to find because their colors and apparent sizes are similar
to those of more luminous, and far more numerous, background galaxies.
By far the best known set of dwarfs is that in the Local Group
\citep{mateo98a,grebel03a,dolphin05a}, where the even the lowest mass
dwarfs can be detected as over-densities in stellar counts
\citep{willman05a, zucker04a}.  The dwarf galaxy population has also
been well studied in nearby clusters, such as Virgo and Fornax, where
the number of dwarfs is large relative the background population
\citep{binggeli85b,ferguson91a} or in other environments where dwarfs
are expected, eg.~as satellites to bright galaxies
\citep{zaritsky97a,gutierrez02a}.  Assembling samples of dwarfs in
more distant and less dense environments has been challenging.  Using
criteria such as surface brightness and irregularity, several authors
have assembled large catalogs of nearby dwarf galaxies
\citep{impey96a, schombert01a, hunter04a,karachentsev04a}.

Addressing the above issues requires a large sample of galaxies across
a range of environments, with minimal selection effects.  Deep,
wide-field optical spectroscopic surveys, such as the Sloan Digital
Sky Survey (SDSS), provide a large, homogeneous sample of dwarf
galaxies covering a wide range of galactic properties and
environments.  Concurrently, blind HI surveys such as the HI Parkes
All Sky Survey (HIPASS; \citealt{meyer04a}) and the upcoming Arecibo
Legacy Fast ALFA Survey (ALFALFA; \citealt{giovanelli05a}) will
provide similarly large samples, with a very complementary set of
selection effects.

Unlike present-day massive disk galaxies, which have used up much of
their gas reservoirs \citep{read05a}, the gaseous component of dwarf
galaxies cannot be ignored \citep{schombert01a,vanzee01a}.  Indeed, we
show that at the lowest luminosities dwarf galaxies are {\it mostly}
neutral gas.  A galaxy's gas content reflects the
efficiency of star formation modified by processes which add or
removal gas from the system such as gas infall or ram pressure
stripping.  The goal of this paper is to combine the optical
properties, dynamics and neutral gas content of low mass dwarf
galaxies in order to better understand the processes important in
dwarf galaxy evolution.

We present neutral hydrogen (HI) observations for \ngal\ low
luminosity dwarf galaxies selected with uniform and well-understood
criteria from the SDSS.  We describe the catalog from which these
dwarfs were selected in \S\,\ref{ssec_sample}, estimate the
environment of each object in \S\,\ref{ssec_environ} and discuss the HI
observations in \S\,\ref{ssec_redux_HI}.  We investigate correlations
between the gas fraction, color, star formation rate and environment
of this sample in \S\,\ref{ssec_fgas} and the Tully-Fisher
relationship in \S\,\ref{ssec_tf}.  When necessary, we assume a Hubble
constant of H$_{0} = 70 h_{70}$\,Mpc \kms\ where $h_{70} = 1$.

\section{Data}
\subsection{The Sample}\label{ssec_sample}

We selected dwarf galaxies with magnitude $M_r -5 \log_{10} h_{70} >
-16$ for HI observations from the low luminosity galaxy catalog of
\citet{blanton04b}, a subsample of the New York University Value-Added
Galaxy Catalog \citep[NYU-VAGC;][]{blanton05a} based on the SDSS
\citep{york00a} Data Release 2 \citep[DR2;][]{abazajian04a}.  The
\citet{blanton04b} catalog represents a significant improvement over
simply selecting galaxies from the SDSS catalog, which is not
optimized for nearby, low surface brightness galaxies.  It is a
flux-limited sample selected with an optical apparent magnitude limit
of $m_r \sim 17.8$.  For each galaxy, the catalog provides the SDSS
redshift, multi-band photometry, structural measurements (in
particular, two-dimensional \Sersic\ fits as described in
\citeauthor{blanton04b}) and emission line measurements
\citep{tremonti04a}.  Distances are estimated based on a model of the
local velocity field \citep{willick97a}.  Distance errors have been
folded into error estimations of all distance-dependent quantities
such as absolute magnitude and HI mass.

There are two selection effects of the \citet{blanton04b} catalog
relevant to this analysis: (1) The SDSS flux limit at $m_r \sim 17.8$
restricts the detection of galaxies with $M_r - 5 \log_{10} h_{70} =
-16$ to within 60\,Mpc in distance (20\,Mpc for the least luminous
galaxies, $M_r - 5 \log_{10} h_{70} = -13.5$).  The catalog therefore
does not span the full range of galaxy environments. While there are
plenty of voids and groups in this volume, there are few very dense
regions. For example, given the angular limits of the SDSS DR2, we see
only the fringes of the Virgo Cluster.  (2) The catalog does not
include low surface brightness galaxies.  As shown in
\citet{blanton04b}, the SDSS completeness as a function of half-light
surface brightness drops below 50\% at $\mu_{50, r} \sim 23.5$ mag
arcsec$^{-2}$.  In the absolute magnitude range of interest, we would
miss half of the dwarf galaxies in the Local Group \citep{mateo98a}
given this surface brightness cutoff.  According to the definition of
\citet{mcgaugh96a}, 10\% of our sample galaxies are considered Low
Surface Brightness (LSB) with $\mu_o^B \ge 22.7$.  With the exception
of these two selection effects, the Blanton et al.~spectroscopic
catalog is not selected by galaxy property in any particular way.  For
example, galaxies have not been selected for or against irregular
looking morphologies.

We present here HI observations for \ngal\ dwarf galaxies in the
absolute magnitude range $-13.5 > M_r - 5 \log_{10} h_{70}> -16$
selected from the \citet{blanton04b} catalog.  These data are part of
a larger, on-going survey to obtain HI line-widths and H$\alpha$
rotation curves for a significant fraction of galaxies in this catalog with
$M_r - 5 \log_{10} h_{70} > -16$.  The complete catalog contains 1034
galaxies at these luminosities.  The \ngal\ galaxies presented here are
representative of the full catalog.

\subsection{Environment and the Nearest Neighbor Distances }\label{ssec_environ}

A goal of this paper is to understand the effects of environment on
the properties of dwarf galaxies.  To quantify the environment, we
calculate for each dwarf galaxy the projected distance to its nearest
luminous neighbor.  We cannot rely on the SDSS alone to determine this
distance for several reasons.  First, the angular distances between
nearest neighbor galaxies can be large for this nearby sample--- for
example, searching a 1\,Mpc region around a galaxy 30\,Mpc away
corresponds to 2 degrees on the sky.  Many of our dwarf galaxies are
on the SDSS Southern stripes, which are only 2.5 degrees wide.  In
addition, because the SDSS reduction software is not optimized for
large extended objects, and fails to process them correctly, the SDSS
catalog does not contain many of the bright galaxies within 30\,Mpc.
Thus, to calculate the environments of our dwarf galaxy sample, we
need a supplemental catalog that extends beyond the SDSS area and
contains the brightest galaxies.

Both of these considerations drive us to use the The Third Reference
Catalog of Galaxies (RC3; \citealt{devaucouleurs91a}), which is a
nearly complete catalog of nearby luminous galaxies. To determine
environments for our dwarf galaxies, we determine the distance to its
nearest ``luminous'' neighbor.  We define ``luminous'' galaxies as
brighter than $M_r - 5 \log_{10} h_{70} < -19$ or have circular
velocities greater than $V_c > 100$\kms\ \citep{haynes99a}.  While we
would like to include galaxies of all sizes in our environment
estimate, the availability of a complete comparison catalog restricts
our search to luminous neighbors.  From the $B$ and $V$ photometry
listed in RC3, we infer $M_r$ for each RC3 galaxy using the {\tt
  kcorrect} product (see \S\,\ref{ssec_fgas}) .  For galaxies which
have the relevant entries listed, we call galaxies luminous if $M_r- 5
\log_{10} h_{70} < -19$.  While this absolute magnitude is set by the
catalog limits, it roughly corresponds to the dividing line between
normal and dwarf galaxies.  For RC3 galaxies which do not have the
relevant entries, but do have HI data listed, we call them luminous if
$W_{20} > 200$ km s$^{-1}$ ($W_{20}$ is twice the maximum circular
velocity of the HI gas, see \S\,\ref{ssec_redux_HI}).  Finally, there
are some galaxies with neither HI data nor optical photometry listed
in RC3. For this small set, we extract the ``magnitude'' from the NASA
Extragalactic Database (NED), which empirically is very similar to the
$B$ band RC3 magnitude, and guess $M_r$ based on that magnitude.
Additionally, we update the coordinates in RC3 using the NED
coordinates for each of the catalog objects.  This set of bright
galaxies is not perfectly uniform, but is suitable for our purposes.

To determine the nearest neighbor distance, we ask whether there is a
luminous RC3 neighbor within a projected distance of 2.5 $h_{70}^{-1}$
Mpc and 300\kms\ in redshift for each dwarf galaxy in our sample.  We
choose these limits because \citet{blanton04c} have shown that
environmental affects are strongest inside this scale.  About 20\% of
our galaxies have no such neighbor; for these objects, we set the
nearest neighbor distance to 2.5\,Mpc. Using the SDSS Catalog Archive
Server\footnote{\tt http://cas.sdss.org}, we have looked at wide field
mosaic images around all of our objects and checked by eye that the
RC3 catalog object is indeed the nearest bright neighbor. This check
is not perfect due to the edge effects of SDSS, but provides a sanity
check on our procedure.

\subsection{HI Observations}\label{ssec_redux_HI}

\begin{figure}
\begin{center}
\includegraphics[angle=0,scale=0.70]{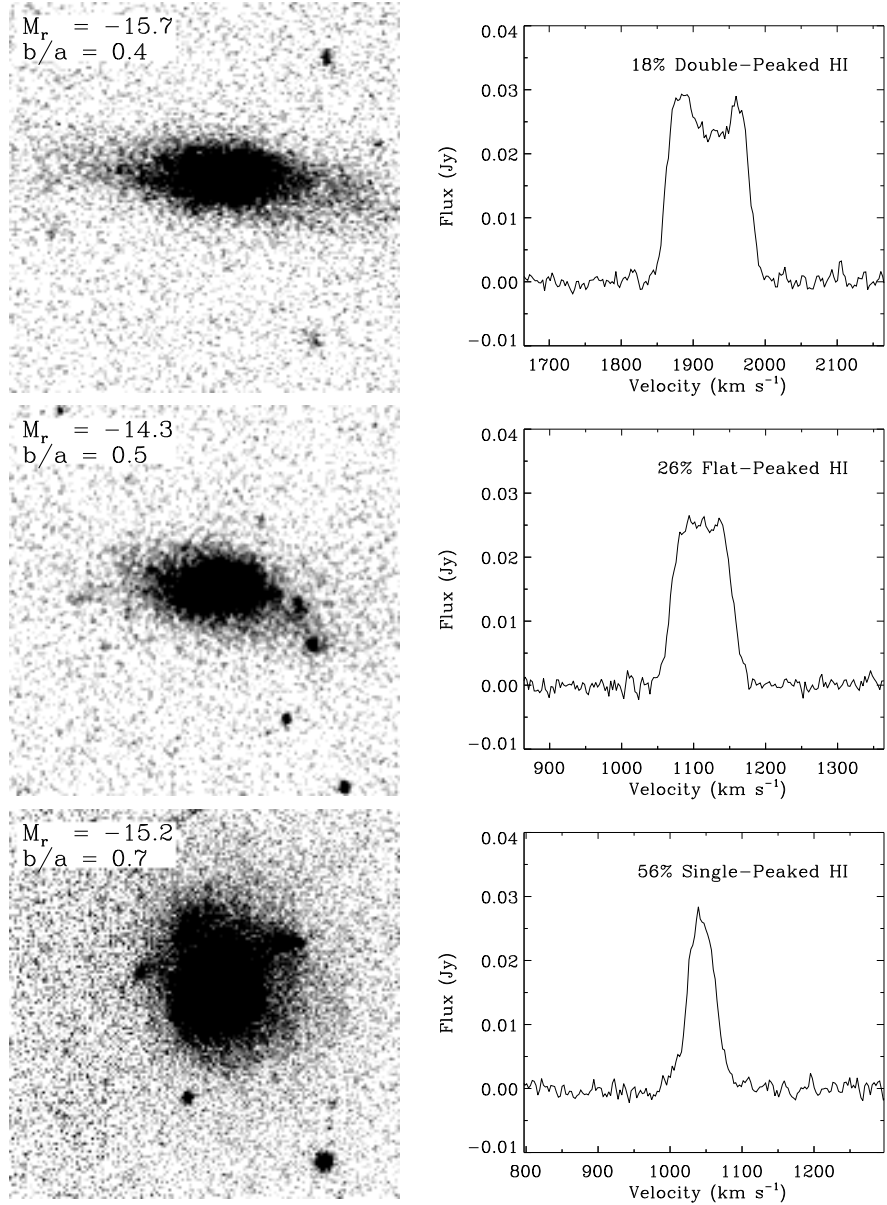}
\epsscale{0.8}
\end{center}
\figcaption{SDSS $r$-band images and HI line-widths for three
representative dwarf galaxies.  The images are $1'$ on a side.  We
list the absolute $r$-band magnitude and optically measured axis ratio
($b/a$) for each galaxy.  Of the dwarf galaxy sample presented here,
18\% have double-peaked HI profiles similar to the top panel, 26\%
have flat-topped HI profiles, and 56\% are single peaked profiles.  If
we consider only edge-on galaxies ($b/a < 0.4$), the percentage of
double-peaked profiles increases to 30\%.  The distribution of HI
profile shapes suggests that the majority of dwarf galaxies in our
sample have regular rotation fields.
\label{fig_imgprof}}
\end{figure}

We obtained HI radio observations with the Arecibo 305-m telescope and
the Green Bank 100-m Telescope (GBT) in July -- August 2005 and March
-- April 2006.  At Arecibo, we used the L-wide receiver in 9-level
sampling mode with 1024 channels and a 12.5\,MHz bandwidth.  At GBT,
we used the L-band with 8192 channels and a bandwidth of 12.5\,MHz.
The Arecibo configuration yields a velocity resolution of 2.5\kms, the
GBT configuration yields a resolution of 0.3\kms.  Standard ON/OFF
position switching was used during both the Arecibo and GBT
observations.  The total on-source integration times were on average
10~minutes for the Arecibo data and 20~minutes for the GBT data.  The
average half-light radius of the dwarf galaxies in our sample is $\sim
8''$ and they should be completely contained within the radio beamsize
of $3'$ and $9'$ for Arecibo and GBT, respectively. We have searched
the beam coverage for nearby contaminating galaxies and have removed
the few cases in which a nearby galaxy is contained in the beam and is
within 100\kms\ of the targeted galaxy. Optical images and HI profiles
for three representative galaxies are shown in
Figure~\ref{fig_imgprof}.  Our observations yielded strong HI
detections for \ngalnoHI\ of \ngal\ galaxies.  The \ngaldiff\ galaxies
which did not yield HI detections are represented by upper limits in
the analysis of \S\,\ref{ssec_fgas} and are excluded from the
Tully-Fisher analysis in \S\,\ref{ssec_tf}, since we cannot estimate
HI line-widths for these galaxies.

We compute the 20\% HI line-width (W20) by finding the peak HI flux
within 150\kms\ of the optical radial velocity of each galaxy and
computing the difference between the nearest points having 20\% of the
peak flux.  The integrated HI flux is calculated by expanding the W20
values by 20\kms\ on a side and integrating the flux in this region.
Errors bars on the line widths and integrated fluxes were computed
using a Monte Carlo bootstrap method: noise was added to the stacked
one-dimensional radio spectra based on the observed variance in the
baseline and the observed quantities measured.  Error bars on the
line-width and integrated flux were calculated from the scatter in the
mean quantities recovered form the Monte Carlo simulations.  We
convert the HI integrated flux into HI mass based on the
optically-thin approximation: $M_{\rm HI} = 2.356\times 10^{5} D^2
F_{\rm HI}$ where D is the distance in Mpc and $F_{\rm HI}$ is the
integrated HI flux in Jy km s$^{-1}$ \citep[Eqn.~9;][]{haynes84b}.  We
note that this standard HI mass conversion is a lower limit to the
true HI mass if the galaxy contains optically thick HI.  Error bars on
the HI mass include the HI flux error and the error in the distance.

\section{Results}\label{sec_results}

\begin{figure}
\includegraphics[angle=0,scale=0.8]{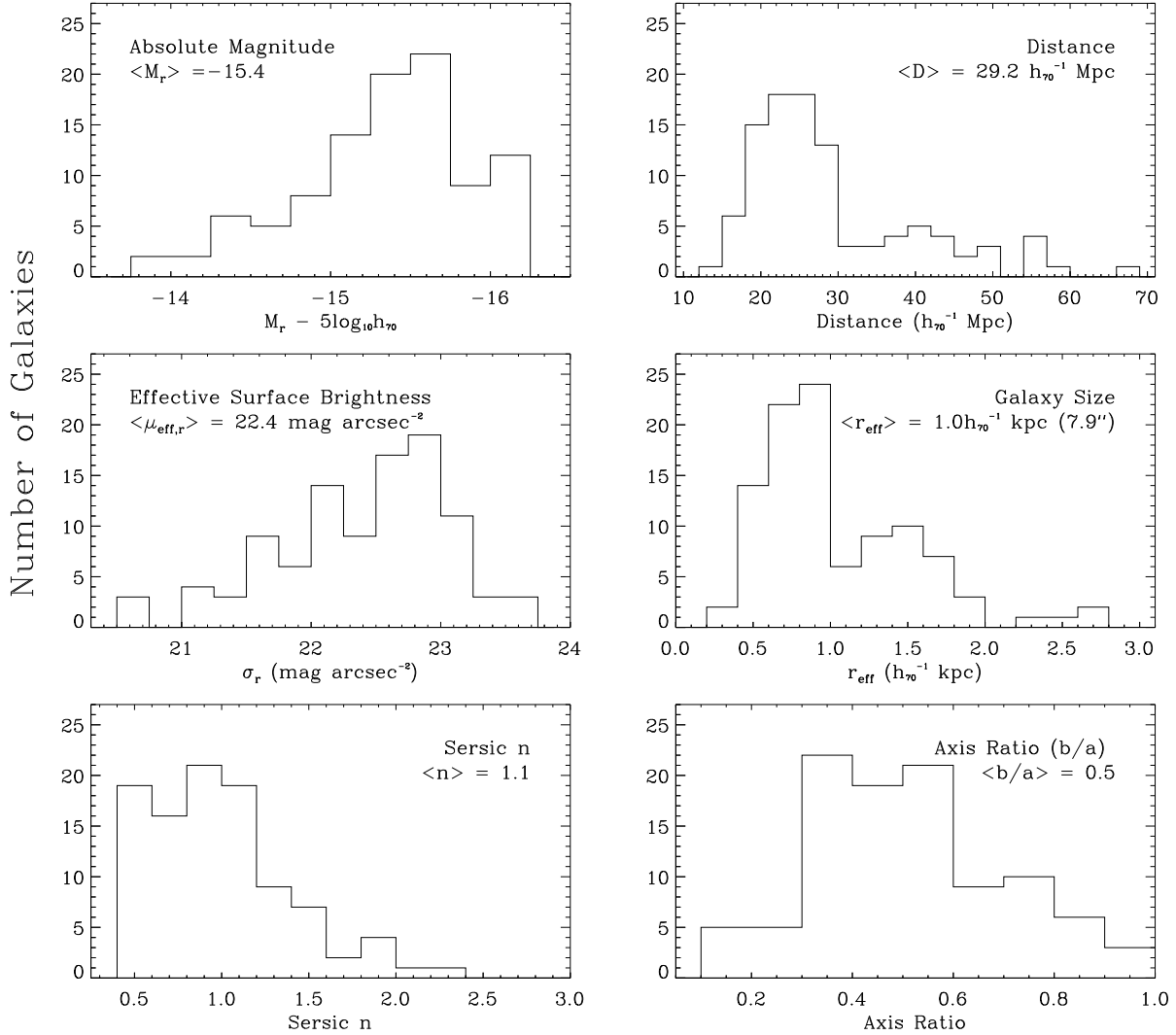}
\epsscale{1.0}
\figcaption{The distribution ({\it top left to bottom right}) of
absolute magnitude, distance, effective surface brightness, effective
half-light radius, \Sersic\ index and axis ratio for our dwarf galaxy
sample. All photometric parameters are measured from the SDSS $r$-band
images.  Distances are estimated based on a model of the local
velocity field \citep{willick97a}. In each panel, we list the
average value of each quantity for the \ngal\ dwarf galaxies in this
sample.\label{fig_hist}}
\end{figure}

The \ngal\ dwarf galaxies presented in this paper represent the
largest homogeneous sample of galaxies fainter than $M_r -5\log_{10}
h_{70} > -16$ with well-measured HI and optical properties.
Properties of the sample are listed in Table~1; the distribution of
these properties is shown in Figure~\ref{fig_hist}.  The average
absolute magnitude, corrected for Galactic extinction
\citep{schlegel98a}, of our sample is $\langle M_r \rangle = -15.4$ at
an average distance of 30\,$h_{70}^{-1}$Mpc.  Galaxies in the sample
tend to be blue, $\langle g-r \rangle$ = 0.3, reflecting the fact that
our sample does not probe the densest environments where red dwarf
ellipticals tend to reside.  The galaxies have an average effective
surface brightness of $\langle \mu_{r,\rm eff} \rangle= 22.4$ mag
arcsec$^{-2}$, and exponential surface brightness profiles as measured
by the \Sersic~index ($\langle n_{\rm Sersic} \rangle = 1.1$).  The
average effective half-light $r$-band radius is $\langle r_{{\rm
    eff},r} \rangle = 8'' = 1.0$\,$h_{70}^{-1}$\,kpc.  These galaxies
exist in environments ranging from dense groups to truly isolated
objects: about 15\% of the sample are satellites of luminous host
galaxies, defined as being within 200 $h_{70}^{-1}$\,kpc projected
distance and 300 km s$^{-1}$ in velocity of a galaxies with $M_r
-5\log_{10} h_{70}< -19$ (see \S\,\ref{ssec_environ}).  Meanwhile,
about 65\% of the galaxies are further than 500 $h_{70}^{-1}$ kpc of
any such host galaxy --- they are very isolated galaxies.

The dwarf galaxies in our sample were not selected with respect to
morphology or gas content.  The sample consists of $\sim10$\% gas-poor
dwarf galaxies commonly known as ``dwarf spheroidal'' or ``dwarf
elliptical'' galaxies.  The majority of our sample consists of
gas-rich dwarfs commonly known as ``dwarf irregular'' galaxies.  This
latter term is commonly misinterpreted to mean that gas-rich dwarf
galaxies have disorganized morphology or turbulent kinematics.
Historically, the term ``irregular'' indicates that the HII regions in
a galaxy do not follow a regular spiral arm pattern as observed in
normal massive disk galaxies \citep{sandage75a,skillman96a}.  Indeed,
our sample dwarf galaxies do not show any evidence of spiral
structure, based on the visual inspection of unsharp-masked SDSS
$r$-band images.  However, our dwarfs do have well-defined optical
centers and regular morphology (Figure~\ref{fig_imgprof}) as has been
noted by previous studies of dwarf galaxies in this regime
\citep{matthews02a}.  Our HI radio profile shapes suggest that our
dwarf galaxies have regular rotation fields: 18\% of the HI profiles
have distinct double-peaked shapes implying a coherent rotation field;
this fraction increases to 30\% if we limit the sample to edge-on
galaxies ($b/a < 0.4 $) only. Thus while the classification ``dwarf
irregular'' is correct for the majority of our sample galaxies, due to
its common misinterpretation, we simply refer to all the galaxies in
this sample simply as low luminosity dwarf galaxies.

\subsection{HI Gas Fractions}\label{ssec_fgas}

We first investigate the ratio of gas mass to total baryonic mass in
our dwarf galaxy sample. We define the gas fraction as $f_{\rm gas} =
M_{\rm gas} / (M_{\rm gas} + M_{\rm stellar})$.  The denominator of
this expression is the total baryonic mass defined as the sum of the
gas and stellar mass.  We take $M_{\rm gas} = 1.4 M_{\rm HI}$, where
$M_{\rm HI}$ is calculated in \S\,\ref{ssec_redux_HI} and the
multiplicative factor takes in account the presence of helium and
metals.  We do not include a correction term for molecular hydrogen
(H$_2$) as the H$_2/$HI ratio is much lower in dwarf galaxies 
compared to luminous spirals \citep{leroy05a,taylor98a}.  This
deficiency holds even after accounting for a strong metallicity
dependence of the H$_2$ tracer molecule CO \citep{garnett02a}.  The
stellar mass, $M_{\rm stellar}$, is based on the optical SDSS $i$-band
magnitude and $(g-r)$ colors using the mass-to-light ratios determined
from the galaxy evolution models of \citet{bell03a} using the
\citet{kroupa93a} initial mass function (that is, we subtract 0.15 dex
from Bell et al.~Table~7 calculated for a 'diet' Salpeter initial mass
function).  We choose this combination of magnitude and color to
minimize contamination from emission lines and thus probe the ``true''
stellar continuum \citep{west05a}.  We have calculated the stellar
mass based on the broad-band fluxes using the {\tt kcorrect}
product\footnote{{\tt http://cosmo.nyu.edu/blanton/kcorrect};
  \citet{blanton03b}} and based on the spectra published by
\citet{kauffmann03a}.  Both these methods produce somewhat lower
stellar mass ($\sim 30\%$) than predicted by Bell et al.  We stress
that the mass-to-light ratios are very sensitive to the assumed
initial mass function.  We note that the gas fraction does not depend
on the distance to individual galaxies.

\begin{figure}
\includegraphics[angle=0,scale=0.7]{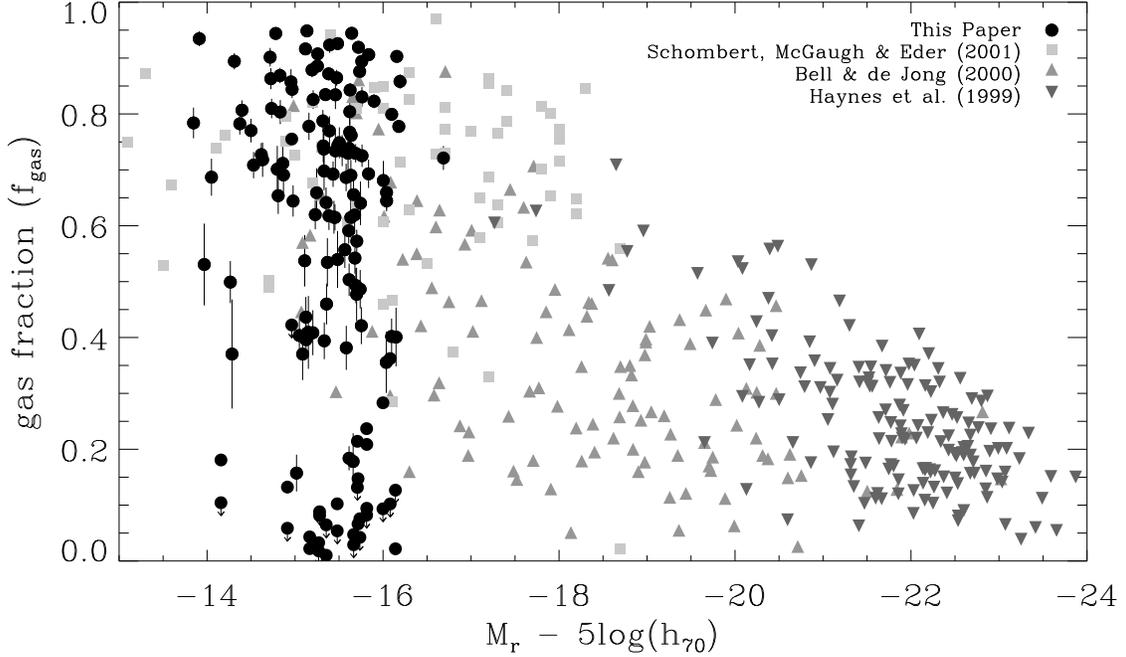}
\figcaption{The gas fraction, $f_{\rm gas} = M_{\rm gas} / (M_{\rm
gas} + M_{\rm stellar}$), plotted as a function of absolute $r$-band
magnitude.  The majority of dwarf galaxies in this paper ({\it black
circles}) are dominated by the gas mass, rather than stellar mass,
with gas fractions as high as \maxfgas\%.  Galaxies which were not detected
in HI are plotted as upper limits.  For comparison, we plot data from
\citet[][{\it squares}]{schombert01a}, \citet[][{\it upward
triangles}]{bell00a} and \citet[][{\it downward
triangles}]{haynes99a}.  We infer from this plot that, compared to
more luminous galaxies, dwarf galaxies are far less efficient at
turning gas into stars over their lifetimes.\label{fig_gas}}
\end{figure}

The average gas fraction for our sample is $\langle f_{\rm gas}\rangle
= $\,\fgas, with gas fractions as high as \maxfgas\%.  Thus, for the
majority of the dwarf galaxies in our sample, the baryonic mass is
dominated by gas, rather than stars.  In Figure~\ref{fig_gas} we plot
gas fraction as a function of absolute $r$-band magnitude.  The flux
limits of our sample are clearly seen in this figure.  We compare our
results to representative samples from the literature by converting
the literature magnitudes to $r$-band using the photometric
transformations from {\tt kcorrect} (see the web page referenced
above) and recalculate gas fractions using the same methods described
above.  The most comparable sample of galaxies is that of LSB dwarf
galaxies by \citet*{schombert01a}.  The gas fractions in this sample
are comparable to our SDSS (higher surface brightness) sample.  For
more luminous spiral galaxies, the average gas fraction is much lower:
\citet{bell00a} find an average gas fraction of 30\% for a sample of
low-inclination spiral galaxies.  The average gas fraction for the
large field spiral sample of \citet{haynes99a} is 25\%, assuming an
average color $V-I =1.0$.  A clear trend is seen in
Figure~\ref{fig_gas}: less luminous galaxies tend toward higher gas
fractions.  While galaxies exist in the lower left region of the plot
(faint, gas-poor dwarf elliptical or dwarf spheroidal galaxies), they
are rare in the upper right region (based on the HIPASS sample, West
et al.~in prep).  In other words, galaxies fainter than $M_r - 5
\log_{10} h_{70} \sim -16$ span the full range of possible gas
fractions, while there is an upper limit to the gas fractions of
brighter galaxies which depends on absolute magnitude.  This trend has
been noted by several authors \citep{mcgaugh97a,schombert01a,lee02a};
Figure~\ref{fig_gas} extends this to significantly fainter absolute
magnitudes compared to previous studies.

We interpret the gas fractions in Figure~\ref{fig_gas} as evidence
that, compared to more luminous galaxies, dwarf galaxies are far less
efficient at turning gas into stars over their lifetimes.  This is
likely related to lower on average gas surface densities in lower
luminosity systems.  \citet{kennicutt98a}, \citet{martin01a} and
references therein have shown for luminous galaxies that star
formation does not occur below a critical gas density ($\sim 5\Msun
{\rm pc}^{-2}$).  \citet{hunter98a} confirm that this limit holds for
low mass systems, although the critical density is slightly lower
($\sim 3\Msun {\rm pc}^{-2}$), due in part to the lack of spiral arms
and solid-body kinematics in the inner regions of dwarfs.  We cannot
precisely measure the average gas density in this sample because our
HI single-dish observations do not provide an estimate of the HI
radius.  However, we can estimate the HI radius by assuming it scales
(roughly) with the optical radius.  We adopt an average HI-to-optical
radius scaling calculated by \citet{begum06a} of $r_{\rm HI} / r_{\rm
H_{\alpha}} = 2.5$.  The average gas density of our sample is $1.5
\Msun {\rm pc}^{-2}$ with more than a factor of two scatter.  Thus,
the majority of galaxies in this sample are at, or below, the critical
gas density required for star formation.  If the threshold described
in \citet{kennicutt98a} exists, the star formation in these dwarf
galaxies implies that the local density of gas in the regions of star
formation significantly exceeds the global density averaged over the
whole galaxy.

\begin{figure}
\includegraphics[angle=0,scale=0.75]{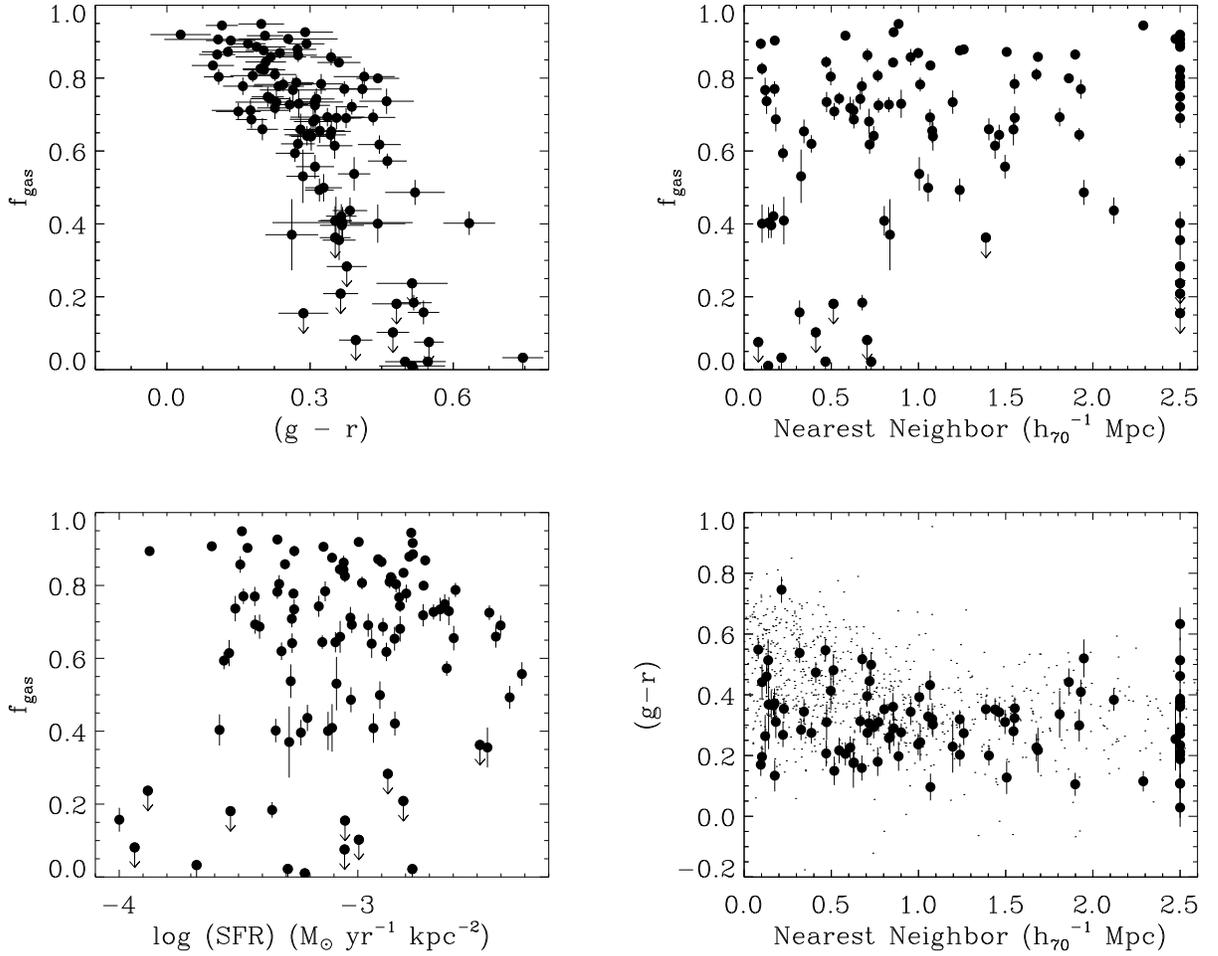}
\figcaption{{\it Top Left:\/} The gas fraction, $f_{\rm gas}$, plotted
  as a function of $g-r$ color.  Assuming the $g-r$ color traces
  recent star-formation activity, the fraction of star-formation
  occurring in the recent past is higher for the gas-rich galaxies in
  our sample. {\it Bottom Left:\/} In contrast, the current star
  formation rate per unit area, as measured by the SDSS $u$-band flux,
  is not correlated with gas fraction.  {\it Top Right\/}: The gas
  fraction plotted against the projected distance to the nearest
  luminous neighbor.  Galaxies with gas fractions $f_{\rm gas} < 0.4$
  are preferentially found within 0.5\,Mpc of a luminous neighbor
  galaxy. The gas fractions of galaxies with $f_{\rm gas} > 0.4$ are
  uncorrelated with nearest neighbor distance. {\it Bottom Right:\/}
  We also compare the $g-r$ color and nearest neighbor distance
  distribution of our sample to that of the parent SDSS dwarf galaxy
  catalog ({\it small black symbols}).
\label{fig_sfr}}
\end{figure}

In the top right panel of Figure~\ref{fig_sfr}, we show that the gas
fraction is a strong function of $g-r$ color in the sense that bluer
galaxies in our sample are gas-rich. We assume that the spread in
$g-r$ color represents a spread in time-averaged star formation over
the past 1\,Gyr, as opposed to a spread in metallicity, for two
reasons. First, the bluest colors in our sample cannot be explained by
old, metal poor stellar population synthesis models. Second, the
majority of galaxies in our sample are currently forming stars (as
indicated by H$\alpha$ emission in the SDSS spectra) which dominates
their broad-band colors \citep{bruzual03a}.  Thus, in the top left of
Figure~\ref{fig_sfr}, the gas-rich galaxies have had a higher fraction
of their star-formation in the recent past.  In contrast, the current
star formation rate per unit area (within the last few million years)
is not correlated with gas fraction (Figure~\ref{fig_sfr}, bottom
left). We calculate this current star formation rate based on the
$u$-band flux using the conversion from \citet{hopkins03a}. For
comparison, we also calculated the star formation rates based on 
 the H$\alpha$ emission line flux in the SDSS
spectra (detected for most of our galaxies), but found that the
resulting rates suffered from the uncertainty of
extrapolating the H$\alpha$ flux from the $3''$ SDSS fiber aperture.
We normalize the total star formation rate by the area over which star
formation is occurring in order to remove any size-dependent
variations within our sample.  The median total star formation rate
per unit area based on the $u$-band for our sample is $SFR =
1\times10^{-3}\Msun$\,yr$^{-1}{\rm kpc}^{-2}$; These rates are
comparable to previous studies of gas-rich galaxies in this luminosity
range \citep{hunter04a,vanzee01a, barazza06a}.  The star formation
rates per unit area are shown in the bottom left of
Figure~\ref{fig_sfr} and are not correlated with gas fraction.  The
gas fraction is better correlated with the time-averaged
star-formation rate over the last billion years ($g-r$ color) than
with the recent star-formation rate in the last few million years.
This implies that the star formation rates we are observing now are
different than those in the recent past; the star formation rates
appear to be stochastic.

\begin{figure}
\includegraphics[angle=0,scale=0.85]{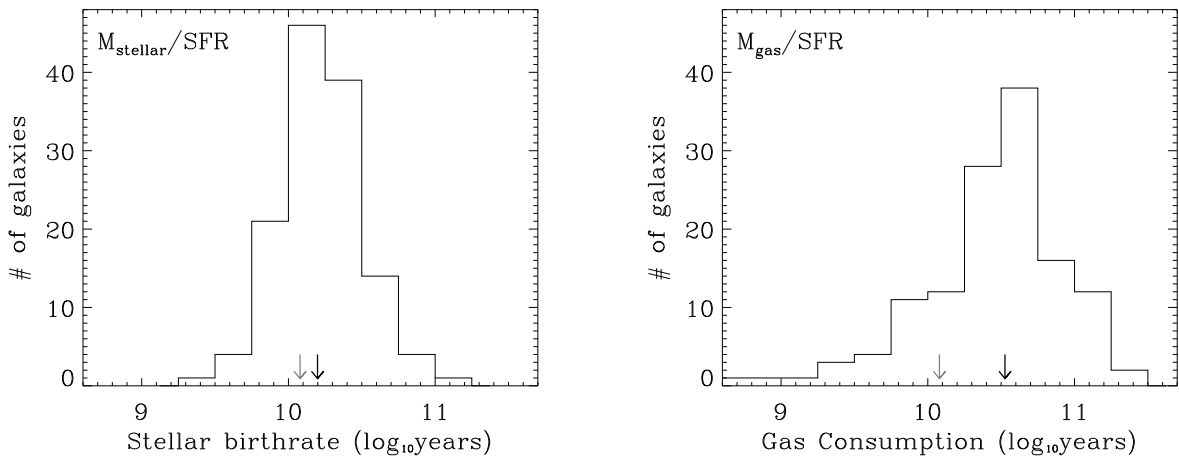}
\figcaption{{\it Left\/}: The stellar birthrate timescale defined as
$M_{\rm stellar} / SFR_{\rm tot}$.  The median birthrate timescale for
this dwarf sample is 15\,Gyr (black arrow).  Assuming a single galaxy
formation time of $\sim$12\,Gyr (grey arrow), roughly half the
galaxies in our sample had lower star formation rates and half have
had higher rates in the past than presently observed.  This suggests
that star formation in dwarfs is stochastic and that the current star
formation rate is not necessarily representative of the time-average
rate. {\it Right:\/} The large gas consumption timescale ($M_{\rm gas}
/ SFR_{\rm tot}$) suggests that the majority of galaxies will be able
to make stars at the current rates for another Hubble
time.\label{fig_hi}}
\end{figure}

We attempt to quantify the stochastic nature of star formation in our
sample by calculating the stellar birthrate timescale.  In the left
panel of Figure~\ref{fig_hi}, we plot the observed stellar mass
divided by the current star formation rate ($M_{\rm stellar} /
SFR_{\rm tot}$).  If the current rate of star formation has been
constant over the lifetime of a galaxy, this quantity would represent
the time needed to form its observed stellar component.  For our
sample, the median timescale to make stars is 15\,Gyrs with a nearly
symmetric scatter to both shorter and longer timescales.  Observations
of the color-magnitude diagrams of dwarf galaxies in the Local Group
suggest that all dwarfs have an ancient stellar population component
\citep{mateo98a,dolphin05a}.  If we assume all the galaxies in our
sample formed at early times (10-12\,Gyr), this implies that roughly
half the sample had lower star formation rates and half had higher
rates in the past than presently observed.  This line of argument
again suggests that star formation in dwarfs fluctuates and that the
current star formation rate is not necessarily representative of the
time-averaged rate.  In the right panel of Figure~\ref{fig_hi} we also
investigate the future of star formation by plotting the gas
consumption timescale ($M_{\rm gas}/SFR_{\rm tot}$).  In this plot we
have not included gas-recycling, which would increase the gas
consumption timescales.  The majority of galaxies in this sample will
be able to make stars, and therefore consume gas, at their present
rates for another Hubble time.

The strongest observed trend in our sample is that between the gas
fraction and $g-r$ color (Figure~\ref{fig_sfr}, top left panel).  This
trend may be due to either internal or external processes.  For
example, a galaxy may use up its gas by forming stars and thus have
redder colors due to a lower star formation rate compared to the
recent past.  Alternatively, a galaxy's gas could be removed in denser
environments and appear redder due to quenching of its star formation.
Since we have quantified the environment of each galaxy (see
\S\,\ref{ssec_environ}), we are able to test whether this and other
trends are due internal or external processes.

The top right panel of Figure~\ref{fig_sfr} suggests that both
internal and external processes are important in controlling the trend
between gas fraction and color.  For gas-rich galaxies in our sample,
with $f_{\rm gas} > 0.4$, the gas fractions are uncorrelated with
nearest neighbor distance.  In contrast, galaxies with low gas
fractions ($f_{\rm gas} < 0.4$) are preferentially found within
0.5\,Mpc of a luminous neighbor galaxy.  Thus, it appears that some
reduction in the gas fraction is related to internal processes such as
normal star-formation, ``primordial'' removal of gas at reionization,
or outflows, however, only environmental effects are very effective at
removing all of the gas from a dwarf galaxy.  This is further
supported by the bottom right panel of Figure~\ref{fig_sfr}. While red
galaxies only exist near luminous neighbors, the colors of the blue
sequence galaxies do not depend on distance to the nearest neighbor.
This result is similar to the relationship between color and
environment, or morphology and environment, seen in more luminous
galaxies (e.g. in \citealt{hubble36a, oemler74a, norberg02a,
  blanton05b}), and the dwarf galaxy population of the Local Group
\citep{grebel03a}.  We conclude that while internal processes are able
to reduce the gas fractions to $f_{\rm gas} \sim 0.4$, external
processes are require to lower the gas fractions (via removal of gas
as suggested below via residuals in the Tully-Fisher relation)
further.

\subsection{The Baryonic Tully-Fisher Relationship}\label{ssec_tf}

The Tully-Fisher relationship relates the mass of a galaxy as inferred
from its luminous component to its dynamical mass, as measured by the
maximum rotational velocity.  The tight coupling between these
components has greatly informed our understanding of galaxy formation
\citep{mo98a, vandenbosch00a, bullock01a, mayer04a}.  Dwarf galaxies
with rotation velocities less than 90\kms\ tend to fall below the
optical Tully-Fisher relationship established by massive galaxies in
the sense that the ratio of stellar to dynamical mass is lower for the
dwarfs \citep{mcgaugh00a,matthews98b}.  This ``break'' has sometimes
been interpreted as evidence that dwarf galaxies have preferentially
lost baryonic mass due to {\it e.g.}~supernovae winds \citep{dekel86a,
  dekel03a}. \citet{mcgaugh00a} first noted that this break is removed
when both the stellar and gaseous components are taken into
account. This baryonic Tully-Fisher relationship has been investigated
by many groups \citep{bell01b, verheijen01b, gurovich04a, mcgaugh05a}.
Understanding the baryonic Tully-Fisher relation for low mass galaxies
should place tight constraints on the role of ram pressure stripping,
tidal stripping, feedback, galactic winds, photoevaporation during
reionization in the evolution of dwarf galaxies.

\begin{figure}
\includegraphics[angle=0,scale=0.85]{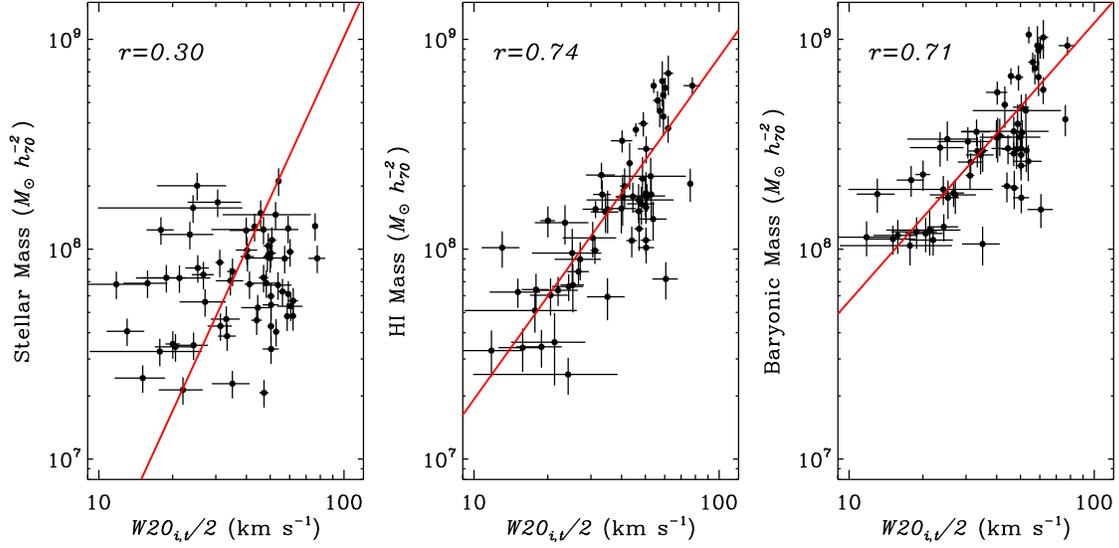}
\figcaption{The stellar mass, HI mass, and total baryonic mass plotted
as a function of rotation velocity as measured by the inclination- and
turbulence-corrected HI half line-widths (W20$_{i,t}$/2).  The solid
line is the (inverse) regression fit described in \S\,\ref{ssec_tf}, whose
parameters are quoted in Table \ref{tffit_table}.  Each panel lists
the correlation coefficient $r$ between the two quantities; the
correlation coefficient is zero for two unrelated quantities and unity
for two perfectly linearly related quantities. The scatter in the
stellar mass relationship ({\it left}) is much larger compared to
the HI ({\it middle}) and baryonic Tully-Fisher relationship ({\it
right}).
\label{fig_tf}}
\vskip 4cm
\end{figure}

\begin{figure}
\includegraphics[angle=0,scale=0.6]{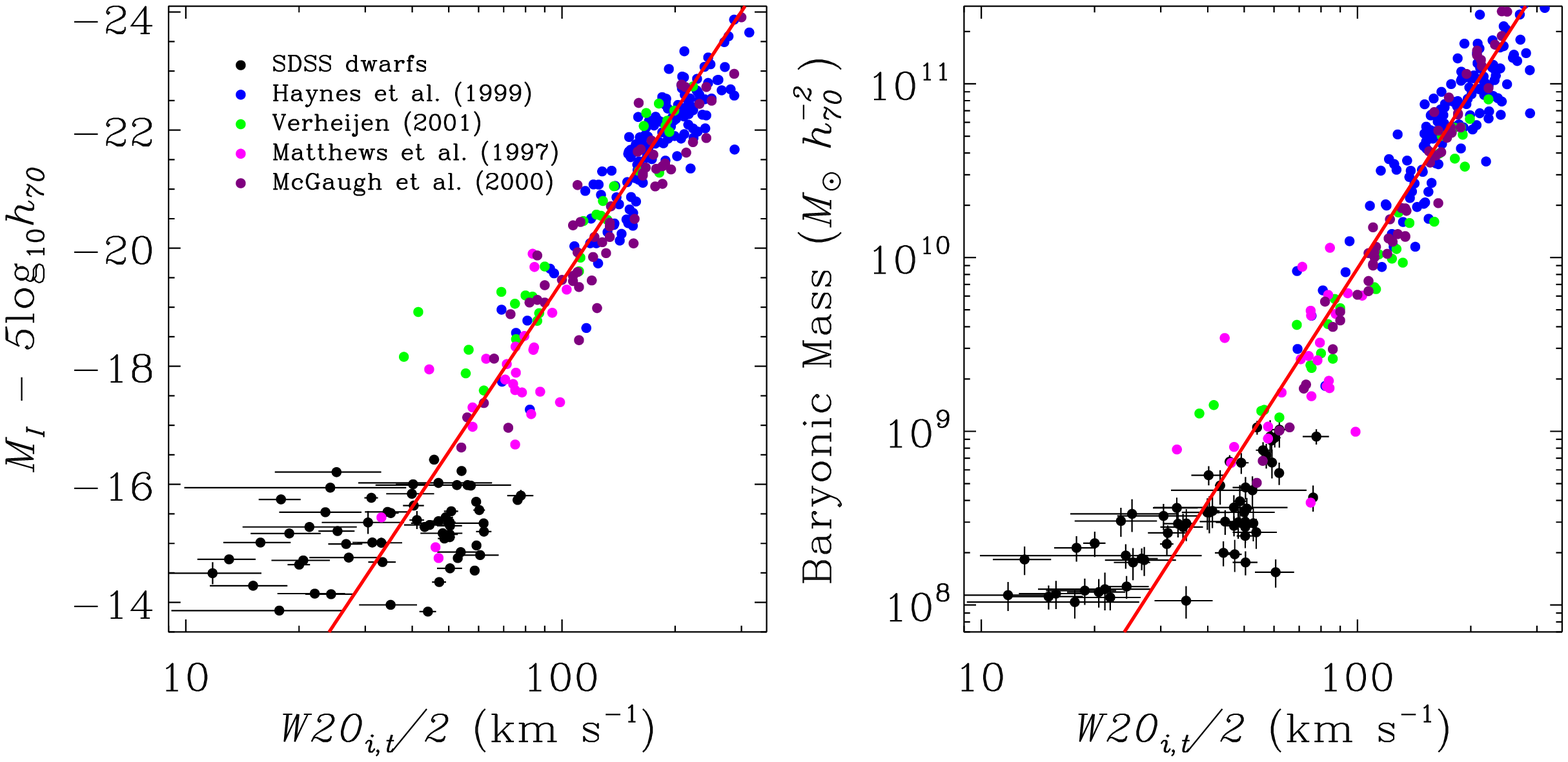}
\figcaption{The optical and baryonic Tully-Fisher relationships:
absolute $I$-band luminosity ({\it left}) and the total baryonic mass
({\it right}) plotted as a function of maximum rotational velocity
(W20$_{i,t}$/2).  We compare our dwarf galaxy sample to literature
data for which HI and optical measurements are available:
\citet{haynes99a}, \citet{verheijen01b}, \citet{matthews98b}, and
\citet{mcgaugh00a}.  The line in each panel is the fit we describe in
the text. The slope in the right-hand panel $b=0.27\pm 0.01$ is
consistent with the CDM prediction $b\sim 0.29$
(\citealt{bullock01a}).
\label{fig_tflit}}
\end{figure}

In Figures~\ref{fig_tf} and~\ref{fig_tflit}, we explore the
Tully-Fisher relationship within our dwarf galaxy sample.  We estimate
the maximum rotational velocities for our galaxies based on the
inclination- and turbulence-corrected HI profile half-width
(W20$_{i,t}$/2).  While we will present resolved optical H$\alpha$
rotation curves for this sample in a forthcoming paper, the HI
line-widths are a more reliable measure of the maximum rotation
velocity in these low mass systems.  This is because the optical
rotation curves of low mass galaxies rarely show a velocity turnover
at the last optically measured data point \citep{matthews02a}, leading
to an underestimate of the true maximum rotation velocity.  Resolved
HI rotation curves also do not always show a velocity turnover
\citep{mcgaugh01a}, however the HI tends to extend 2-3 times beyond
the observed optical radii \citep{begum06a} and thus is a more
reliable probe of the true maximum rotation velocity.

The dwarf galaxies in our sample have a significant rotation component
(Figure~\ref{fig_imgprof}).  We correct the observed HI line-widths
for line broadening due to turbulent velocity dispersion and
inclination using the formula first proposed by \citet{bottinelli83a}:
\begin{equation}
\mathrm{W20}_{\it i,t} = \frac{\mathrm{W20}-\mathrm{W20}_{\it t}}{\sin{\it i}}
\end{equation}
Where W20 is the observed HI line-width, W20$_t$ is the turbulent
velocity correction term and $i$ is the inclination angle inferred
from the optical images.  We confirm the validity of a linear
turbulence correction based on modeling the integrated velocity
profiles of simulated galaxies constructed from \citet{hernquist93a}
model disk galaxies.  For nearby dwarf galaxies with rotation
velocities similar to our sample, \citet{begum06a} have measured
velocity dispersion in the gas component of $\sigma_{\rm los} =
8$\kms\, based on the average line-of-sight velocity dispersion
measured from 2D velocity maps.  Using the Begum et al.~value, this
results in a turbulence correction of W20$_t = 16$\kms, which we use
here.  Altering this turbulence correction (say to 25 \kms) does not
change our results below.

Using the approach of \citet{haynes84b}, we infer $\sin i$ from the
$b/a$ axis ratio based on two-dimensional \Sersic~fitting of the
$r$-band SDSS images. In particular, we assume the disks to be
intrinsically oblate and axisymmetric, with a three-dimensional axis
ratio of $q_0$, and take:
\begin{equation}
\sin i = \sqrt{\frac{1- (b/a)^2}{1-q_0^2} },
\end{equation}    
with $q_0=0.19$. Again, changing the particular value of $q_0$ (say to
0.3) again does not change our results significantly.

Figure~\ref{fig_tf} shows the stellar, HI and baryonic Tully-Fisher
relationships for our dwarf galaxy sample.  In the three panels of
this figure, we linearly regress to fit a function of the form:
\begin{equation}
\label{tffit_eqn}
\log_{10}\left(\frac{\mathrm{W20}_{\it i,t}/2}{50 \mathrm{km s}^{-1}}\right) = 
a + b \log_{10} \left(\frac{\mathrm{mass}}{10^8 h_{70}^{-2} M_\odot}
\right)
\end{equation}
where $a$ and $b$ are the fitted intercept and slope of the
Tully-Fisher relationship, respectively. These fits are regressions
with mass as the independent variable. Table
\ref{tffit_table} lists their values and uncertainties (calculated
from bootstrap resampling). In addition, we calculate the correlation
coefficient $r$ between each pair of variables, listed in each panel
of Figure \ref{fig_tf}.  The correlation coefficient is zero for two
unrelated quantities and unity for two perfectly linearly related
quantities \citep[e.g.~][]{lupton93a}.

The stellar Tully-Fisher (Figure~\ref{fig_tf}, right panel) has
significantly more scatter than the either the HI or total baryonic
relationships (Figure~\ref{fig_tf}, middle and left panels).  The
stellar relation has a much lower correlation coefficient ($r\sim 0.3$
versus $r\sim 0.8$) and considerably more scatter about the best fit
regression line, as listed in Table \ref{tffit_table} (about 50\%
scatter compared to about 30\% scatter), than do the HI mass or
baryonic mass TF relationships.  This result is perhaps not
surprising: at low luminosities, stars contribute far less to the
total baryonic mass compared to massive galaxies
(Figure~\ref{fig_gas}).  Thus, for dwarf galaxies of the same total
mass, the amount of gas turned into stars varies between systems.
While the scatter in the baryonic Tully-Fisher relationship is much
smaller compared to the stellar relationship, it does have more
scatter than accounted for by our observational error bars.  The
fitted slope is unchanged if we restrict our sample to only galaxies
whose inclination corrections are small ($b/a < 0.4$) or if we alter
our assumptions about turbulence or the intrinsic axis-ratio.

In Figure~\ref{fig_tflit}, we compare our dwarf galaxy sample to
Tully-Fisher data available in the literature.  We compare only to
literature samples for which HI and optical measurements are
available: \citet{haynes99a}, \citet{verheijen01b},
\citet{matthews98b}, and \citet{mcgaugh00a}.  For the literature
datasets, we recalculate distances based on $H_0 = 70$\,Mpc \kms\ and
a model of the local velocity field \citep{willick97a} (for galaxies
with recession velocities less than 6000\kms).  For the
\citet{haynes99a} data, we assume a constant color of $V-I = 1.0$.  

%
%

In the left panel of Figure~\ref{fig_tflit} we plot the optical
Tully-Fisher relationship in the $I$-band.  \citet{mcgaugh00a} first
noted that low mass dwarf galaxies fall well below the optical
Tully-Fisher (TF) relationship as defined by luminous spiral galaxies
in the sense that, for a given rotation velocity, the dwarfs are
optically under-luminous compared to the expected TF value.
\citeauthor{mcgaugh00a}~also noted that these same galaxies lie on the
same relation defined by luminous galaxies in the baryonic TF
relationship (Figure~\ref{fig_tflit}, right panel).  Again, this is
not surprising: we have shown above that dwarf galaxies have high gas
fractions and the optical luminosity/stellar mass is often only a few
percent of the total baryonic mass.  However, unlike
\citeauthor{mcgaugh00a}, we do not see evidence for a distinct break
in slope of the optical TF occurring at 90\kms.  At low luminosities,
the scatter around the optical TF relation does increase considerably,
but not preferentially to high or low velocities.  

That a single linear fit adequately describes the baryonic TF data
over more than three decades of mass is significant.  Many authors
have suggested that supernovae-driven winds will preferentially remove
baryonic material from galaxies with $v_{\rm max} < 100$ km s$^{-1}$
due to their far smaller gravitational potentials compared to luminous
galaxies \citep{dekel86a, dekel03a}.  Figure~\ref{fig_tflit} suggests
that a critical mass threshold does not exist below which
mass-ejection is efficient down to baryonic masses of $10^8$\Msun\ and
rotational velocities of as low as 20\kms.  If a critical threshold
existed, we would have expected to see a break in the baryonic
Tully-Fisher.

While it might be possible to explain this single relationship via a
baryonic mass-loss rate which varies smoothly with mass, the measured
slope of this relation suggests that this is not the case.  From the
virial theorem, the expected slope of the baryonic TF is $b=0.33$.
The predicted slope from cosmological simulations is $b\sim 0.29$, or
as more commonly quoted $1/b \sim 3.5$ \citep{bullock01a, kravtsov04a}, in
good agreement with our value of $b = 0.27 \pm 0.01$, or $1/b =
3.70\pm0.15$. Thus, we interpret the baryonic TF relationship shown in
Figure~\ref{fig_tflit} as evidence that dwarf galaxies do not
preferentially lose significant baryonic mass compared to luminous
galaxies.

\begin{figure}
\includegraphics[angle=0,scale=0.7]{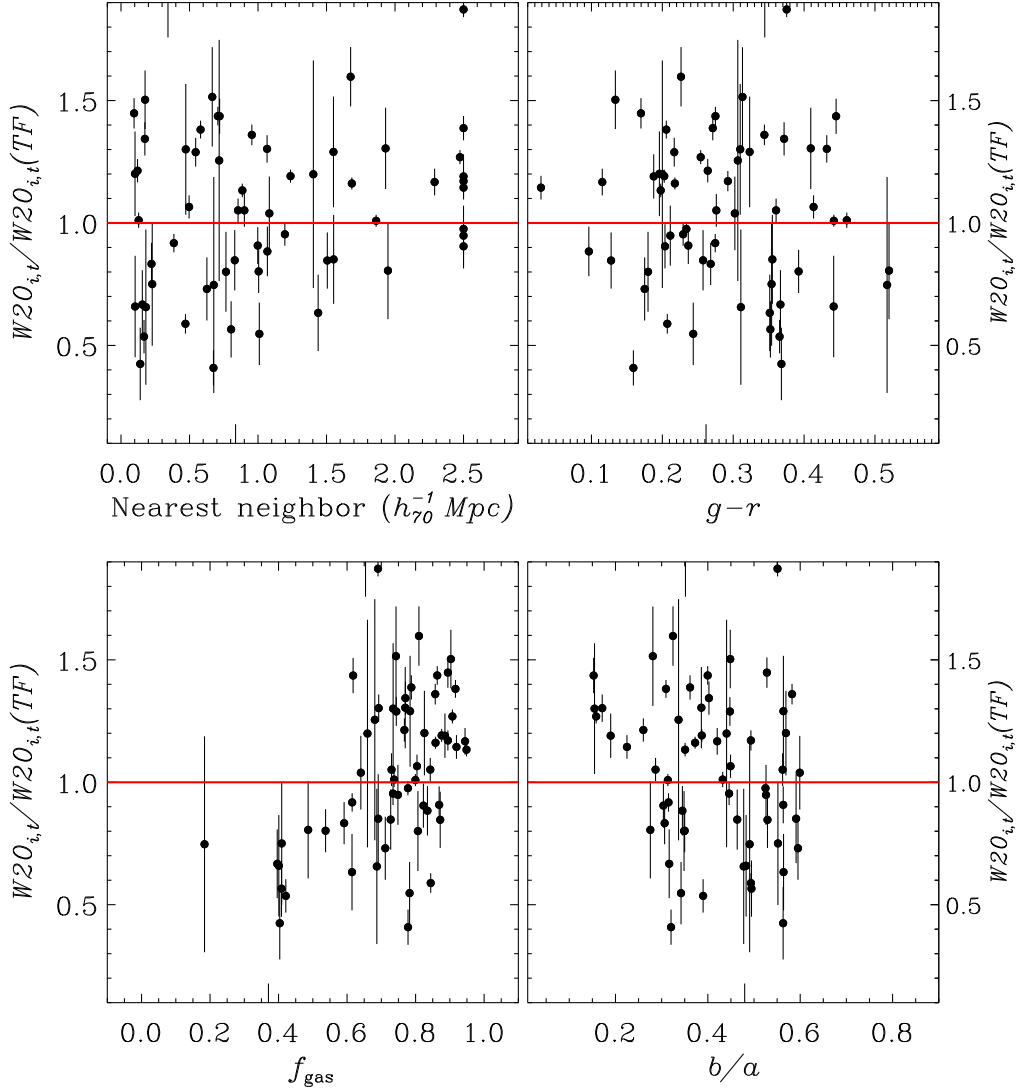}
\vskip -1cm
\figcaption{ Residuals from the baryonic Tully-Fisher
relationship, defined as the ratio $W20_{i,t}/W20_{i,t}(TF)$ of the
actual line width to the expected line width from the Tully-Fisher
relationship. We show this ratio versus nearest neighbor distance,
$g-r$ color, gas fraction $f_{\mathrm{gas}}$ and optical axis ratio
$b/a$. \label{fig_resid}}
\end{figure}

Finally, we consider the residuals from the TF relationship. We define
these residuals in terms of the ratio $W20_{i,t}/W20_{i,t}(TF)$, the
corrected observed HI line-width divided by that expected from the
fitted linear baryonic Tully-Fisher relationship.  This quantity is
defined in the sense that larger $W20_{i,t}/W20_{i,t}(TF)$ values
correspond to lower inferred baryon-to-dark matter ratios.  Figure
\ref{fig_resid} shows this quantity as a function of projected nearest
neighbor distance, color, gas fraction and axis ratio $b/a$.  We find
no correlation between these residuals and the axis-ratio $b/a$,
except that the very flat galaxies ($b/a < 0.25$) never exhibit low
$W20_{i,t}/W20_{i,t}(TF)$ ratios (Figure~\ref{fig_resid}, bottom left
panel).  That no strong trend exists with axis-ratio suggests that the
measured HI line-widths are properly sampling the galactic potential.
Dwarf galaxies with low gas fractions have low
$W20_{i,t}/W20_{i,t}(TF)$ ratios.  This suggests that low gas fraction
dwarf galaxies, which are preferentially found in dense environments,
have {\it higher} baryon-to-dark matter ratios compared to the
majority of dwarfs in our sample.  If, for example, reduced gas
fractions in dense environments were due to interaction-induced star
formation, we would expect these ratios to remain constant.  If this
were due to ram pressure stripping alone one might expect this ratio
to be lower in dense environments: gas stripping would preferentially
remove baryonic mass, reducing the ratio of baryons to dark matter.
There is little correlation with color, indicating that if gas
stripping is important, it is not directly effecting to the
star-formation rates at the center.  Alternatively, it is possible
that this is simply a selection effect and the HI line-widths do not
properly sample the galactic potentials of the lowest gas fraction
galaxies.  As the number of observed galaxies in this mass range
increase, the TF residuals should place significant constraints on
galaxy evolution processes important at these masses.

\section{Conclusions}

We present results from an HI survey of low luminosity dwarf galaxies
in the absolute magnitude range $-13.5 > M_r - 5\log_{10} h_{70} >
-16$ selected from the SDSS.  These data represent the largest
homogeneous sample of low luminosity dwarf galaxies with well-measured
HI and optical properties.  The sample spans a range of total baryonic
masses between M$_{\rm baryonic} = 10^8-10^9$\Msun with measured HI
line-widths between $\langle v_{\rm
  rot}\rangle=\langle$W$20_{i,t}/2\rangle=20 - 80$ \kms.  The dwarf
galaxies in this sample are found in a wide range of environments,
from dense groups to truly isolated galaxies.

For the majority of our dwarf galaxy sample, the total galactic
baryonic mass is dominated by the gas mass rather than stellar mass.
The average gas fraction for our sample is $\langle f_{\rm
  gas}\rangle=$\,\fgas, with gas fractions as high as \maxfgas\%.
This significantly exceeds that of gas-rich galaxies at higher
luminosities.  Thus, dwarf galaxies are inefficient at turning gas
into stars compared to more luminous galaxies.  We find that gas-rich
galaxies have bluer $g-r$ colors compared to gas-poor objects implying
that gas-rich dwarf galaxies have had more recent star formation.  The
majority of galaxies in our sample are currently forming stars based
on observed H$\alpha$ emission in the SDSS spectra.  We find that the
{\it current} rate of star formation per unit area (based on the
total $u$-band flux) is not correlated with gas fraction, confirming previous
results suggesting that the star formation rates in dwarf galaxies are
stochastic \citep{vanzee01a,hunter04a}.  Thus, dwarf galaxies form
stars with a rate that varies over million year time scales, but is
correlated with the gas fraction over longer time periods.

Using the range of environments within our sample, we examine whether
internal or external processes control the gas fractions of dwarf
galaxies.  Trends between galaxy properties, such as color or
morphology, and environment have been well studied for galaxies at
brighter luminosities \citep{norberg02a,blanton05b}, but have been
previously limited to the Local Group and nearby groups at dwarf
luminosities \citep{grebel03a}.  Our sample allows us to quantify
these trends for dwarf galaxies over a much wider range of
environments.  We find that dwarf galaxies with gas fractions $f_{\rm
  gas} < 0.4$, comprising $15\%$ of this sample, are exclusively found
within 0.5\,Mpc of a luminous neighbor galaxy.  Galaxies with gas
fractions $f_{\rm gas} > 0.4$ are found across the full range of
environments contained within this sample.  Thus, while internal
galaxies processes can reduce the gas fractions, external processes
are required to fully remove gas from a dwarf galaxy.

The optical Tully-Fisher relationship within our dwarf galaxy sample
demonstrates considerable scatter.  As previous work has shown for
brighter samples (\citealt{bell01b, verheijen01b, gurovich04a,
  mcgaugh05a}), the baryonic Tully-Fisher relationship within our
sample is much tighter. When we consider the Tully--Fisher
relationship across all absolute magnitudes, we find a logarithmic
slope of $b = 0.27\pm 0.01$ (or $1/b=3.70 \pm 0.15$). These results
are in accord with CDM predictions of $1/b \sim 3.5$ assuming no
preferential loss of baryons as a function of mass
(\citealt{bullock01a}). Thus, our results suggest that processes which
preferentially remove gas from dwarf galaxies are not important for
the majority of galaxies in our sample with maximum rotation
velocities $v_{\rm max} \sim 20 - 80$ \kms\ (roughly corresponding to
total dynamical masses of M$_{\rm total}=10^9-10^{10}$\Msun).

If galaxies are associated one-to-one with $\Lambda$CDM dark matter
halos, the mass-to-light ratios of galaxies must decline as a function
of host halo mass \citep{benson03a}.  The most well-known example of
this problem is the Local Group ``substructure problem''
\citep{klypin99a,moore99a}.  To explain this trend in mass-to-light
ratio, investigators often invoke processes such as ram pressure
stripping or supernova feedback that either removes gas or heats it
sufficiently to prevent star-formation in low mass galaxies (e.g.,
\citealt{dekel86a,dekel03a,mori00a,bullock00a}).  Most semi-analytic
models require significant baryonic mass loss
\citep{cole00a,benson03a,croton06a}, predicting baryon-to-dark matter
ratios for dwarf galaxies well below the ``Universal'' baryon fraction
\citep{read05a}.  Comparing the metallicity of galaxies over a wide
range of baryonic masses, \citet{tremonti04a} found that dwarfs are
metal depleted relative to massive galaxies and argue in favor of
galactic winds efficiently removing enriched material from low mass
galaxies.  However, at the masses of the galaxies in our sample,
detailed models of the physics of gas blow-out and blow-away tend to
disfavor significant gas mass loss, predicting loss of only a few
percent \citep{maclow99a,ferrara00a}.  Our sample cannot rule out
small amounts baryon mass loss from these galaxies, nor scenarios in
which gas remains bound to the galaxy but is blown away from the
region participating in star formation.  This may explain the low
observed metallicities of dwarf galaxies --- enriched gas is removed
from regions of active star formation, resulting in low observed
metallicities, but either remains bound or represents only a few
percent of the total baryonic mass.  Indeed, recent cosmological
simulations which include realistic feedback and chemical evolution
suggest that the metallicity, mass-to-light ratios and other galaxies
properties can be explained without including supernova energy
feedback \citep{tassis06a}.  Our results favor such dwarf galaxy
formation models in which processes such as supernovae winds are not
responsible for significant gas or baryon mass in low mass galaxies.

\acknowledgments 

We thank David Weinberg and Rob Kennicutt for early and
important encouragement on this project.  We had very productive
conversations with many people regarding this paper including Andrey
Kravtsov, Jeremy Darling, Oleg Gnedin, Evan Skillman, Beth Willman,
Jay Gallagher, Ari Maller, Kathryn Johnston, David W.~Hogg, and
St\'ephane Courteau.  We would like to thank Carl Bignell, Toney
Minter, and Karen O'Neil for help with the GBT observations.  We thank
Martin Kirby for creative assistance in naming the project.  M.~G.~is
supported by NASA through Hubble Fellowship grant HF-01159.01-A
awarded by the Space Telescope Science Institute, which is operated by
the Association of Universities for Research in Astronomy. M.~M.~is
supported by NASA NAG5-11669 and NSF AST-0428465.  A.~W.~is supported
by NSF AST-0540567.




\clearpage
\begin{deluxetable}{lccccccccccccccc}
\rotate
\tabletypesize{\tiny}
\tablecaption{Low Mass SDSS Dwarf Galaxy Properties}
\tablewidth{0pt}
\tablehead{
\colhead{Name} &
\colhead{$\alpha$ (J2000)} &
\colhead{$\delta$ (J2000)} &
\colhead{Distance} &
\colhead{m$_r$} &
\colhead{$M_r$} &
\colhead{$g-r$}&
\colhead{$\mu_{{\rm eff},r}$}&
\colhead{$r_{{\rm eff},r}$}&
\colhead{$b/a$}&
\colhead{$M_{\rm stellar}$}&
\colhead{$M_{\rm HI}$}&
\colhead{$\sigma_{M_{\rm HI}}$}&
\colhead{W20/2}&
\colhead{W20$_{i,t}/2$}&
\colhead{$\sigma_{W20_{i,t}/2}$} \\
\colhead{}&
\colhead{(h$\,$:$\,$m$\,$:$\,$s)} &
\colhead{($^\circ\,$:$\,'\,$:$\,''$)} &
\colhead{{\tiny ($h_{70}^{-1}$ Mpc)}} &
\colhead{}&
\colhead{}&
\colhead{}&
\colhead{}&
\colhead{kpc ($''$)}&
\colhead{}&
\colhead{{\tiny ($log M\sun$)}}&
\colhead{{\tiny ($log M\sun$)}}&
\colhead{{\tiny ($log M\sun$)}}&
\colhead{\kms} &
\colhead{\kms} &
\colhead{\kms}
}
\startdata
      386141 &  00 01 03.6   & +14 34 48.6 &   15.3 &   16.72 & $-$15.12 &   0.38 &   20.3 & 0.6 (7.8) & 0.65 &  7.82 &  7.56 &  0.08 &  28.7 &  26.6 &   4.7 \\
      241112 &  00 13 38.7   & +15 40 30.5 &   17.9 &   17.02 & $-$15.26 &   0.25 &   21.5 & 1.9 (21.5) & 0.16 &  7.79 &  8.63 &  0.08 &  71.8 &  63.8 &   1.3 \\
      192963 &  00 31 27.6   & $-$10 40 33.2 &   35.1 &   17.43 & $-$16.16 &   0.13 &   22.5 & 1.9 (11.4) & 0.45 &  7.96 &  8.78 &  0.04 &  85.3 &  85.0 &   6.4 \\
      192971 &  00 32 31.3   & $-$10 41 21.9 &   31.4 &   17.45 & $-$15.90 &   0.20 &   20.7 & 1.0 (6.6) & 0.30 &  8.00 &  8.52 &  0.05 &  52.7 &  46.0 &   4.1 \\
      667743 &  00 36 31.7   & +00 33 47.8 &   40.2 &   17.72 & $-$16.14 &   0.50 &   18.8 & 0.6 (3.1) & 0.56 &  8.43 & . & . & . & . & . \\
      190632 &  00 47 51.9   & $-$11 10 29.6 &   39.3 &   17.70 & $-$16.10 &   0.44 &   19.3 & 0.7 (3.9) & 0.31 &  8.32 &  8.78 &  0.03 &  66.3 &  60.3 &   1.4 \\
      677002 &  00 57 56.6   & +00 52 08.9 &   21.0 &   16.96 & $-$15.50 &   0.21 &   21.3 & 0.5 (4.7) & 0.53 &  7.85 &  8.18 &  0.08 &  43.2 &  40.6 &   4.6 \\
       47936 &  00 58 55.5   & +13 43 15.1 &   48.2 &   17.74 & $-$16.69 &   0.39 &   18.8 & 0.4 (1.5) & 0.80 &  8.51 &  8.77 &  0.06 &  63.9 &  92.0 &  23.7 \\
      677307 &  01 14 20.3   & +00 55 00.0 &   10.7 &   16.69 & $-$14.31 &   0.17 &   22.1 & 0.4 (8.1) & 0.53 &  7.32 &  8.10 &  0.08 &  54.5 &  53.8 &   2.1 \\
      222989 &  01 17 30.5   & $-$09 17 48.4 &   17.6 &   15.91 & $-$16.19 &   0.22 &   20.8 & 1.4 (16.2) & 0.37 &  8.10 &  8.73 &  0.06 &  70.0 &  65.6 &   1.3 \\
      198542 &  01 19 14.3   & $-$09 35 46.3 &   17.3 &   17.35 & $-$14.73 &   0.23 &   21.5 & 0.5 (6.0) & 0.32 &  7.52 &  8.01 &  0.06 &  62.5 &  56.5 &   3.9 \\
      461714 &  01 26 04.7   & +00 18 54.9 &   17.3 &   16.33 & $-$15.73 &   0.20 &   21.4 & 0.9 (11.2) & 0.39 &  7.96 &  8.66 &  0.07 &  67.8 &  63.7 &   1.3 \\
      203452 &  01 27 25.1   & $-$08 41 21.5 &   38.8 &   17.77 & $-$16.00 &   0.38 &   19.0 & 0.7 (3.7) & 0.49 &  8.22 & . & . & . & . & . \\
      203478 &  01 32 22.3   & $-$08 38 02.4 &   38.6 &   17.59 & $-$16.18 &   0.23 &   22.4 & 1.1 (5.7) & 0.52 &  8.17 &  8.57 &  0.03 &  53.3 &  52.3 &   1.4 \\
      198720 &  01 41 39.7   & $-$09 13 03.8 &   16.5 &   16.59 & $-$15.35 &   0.29 &   21.5 & 0.7 (8.4) & 0.72 &  7.86 &  7.97 &  0.07 &  39.0 &  43.7 &   2.4 \\
      191112 &  01 50 54.5   & $-$10 22 10.6 &   16.4 &   17.46 & $-$14.50 &   0.37 &   21.4 & 0.6 (7.3) & 0.40 &  7.66 &  8.04 &  0.07 &  54.7 &  50.1 &   2.3 \\
      225456 &  01 53 01.4   & $-$09 38 21.7 &   15.0 &   17.67 & $-$14.05 &   0.31 &   22.5 & 0.5 (6.2) & 0.48 &  7.51 &  7.71 &  0.08 &  28.8 &  23.3 &   8.8 \\
       48406 &  02 00 02.0   & +12 32 18.0 &   34.5 &   17.53 & $-$16.15 &   0.44 &   21.9 & 1.3 (7.5) & 0.48 &  8.30 &  7.98 &  0.12 &  35.6 &  30.9 &   8.1 \\
      201346 &  02 02 46.9   & $-$08 27 27.1 &   21.0 &   17.07 & $-$15.39 &   0.41 &   22.6 & 1.2 (11.4) & 0.39 &  7.84 &  8.22 &  0.08 &  59.1 &  54.4 &   6.3 \\
      231588 &  02 05 15.8   & $-$09 33 46.7 &   17.9 &   17.28 & $-$14.83 &   0.24 &   20.8 & 0.3 (3.8) & 0.56 &  7.59 &  8.26 &  0.07 &  41.4 &  39.7 &   2.9 \\
      227294 &  02 17 56.5   & $-$08 51 14.2 &   13.5 &   16.53 & $-$14.98 &   0.34 &   21.5 & 0.4 (6.3) & 0.77 &  7.77 &  7.88 &  0.07 &  45.2 &  57.2 &   2.8 \\
      643417 &  02 21 18.9   & $-$00 47 55.1 &   12.8 &   17.23 & $-$14.16 &   0.48 &   21.1 & 0.5 (8.0) & 0.67 &  7.57 & . & . & . & . & . \\
      462731 &  02 52 16.8   & +00 17 41.3 &   14.7 &   17.21 & $-$14.63 &   0.23 &   21.2 & 0.3 (3.6) & 0.99 &  7.55 &  7.81 &  0.08 &  39.0 & 301.9 &  63.0 \\
      201616 &  02 53 46.4   & $-$07 23 43.6 &   13.1 &   16.60 & $-$14.86 &   0.18 &   21.1 & 0.6 (8.8) & 0.60 &  7.53 &  7.78 &  0.08 &  29.8 &  26.6 &   3.7 \\
      467776 &  03 07 15.7   & +00 43 52.1 &   28.7 &   17.60 & $-$15.74 &   0.30 &   20.7 & 0.5 (3.9) & 0.60 &  8.09 &  8.19 &  0.09 &  46.0 &  46.6 &   5.9 \\
      199465 &  03 37 19.4   & $-$06 21 13.2 &   30.5 &   17.50 & $-$15.81 &   0.51 &   21.7 & 1.2 (8.3) & 0.45 &  8.14 & . & . & . & . & . \\
      204350 &  03 38 45.7   & $-$05 38 42.1 &   31.5 &   17.61 & $-$15.81 &   0.36 &   19.1 & 0.5 (3.0) & 0.54 &  8.15 & . & . & . & . & . \\
      161656 &  08 48 18.7   & +01 15 50.4 &   19.3 &   16.58 & $-$15.74 &   0.52 &   21.6 & 1.1 (12.3) & 0.28 &  8.22 &  8.05 &  0.08 &  43.0 &  35.8 &   7.7 \\
      123408 &  09 14 03.2   & +60 14 17.3 &   18.7 &   16.51 & $-$15.75 &   0.37 &   20.4 & 0.6 (6.8) & 0.39 &  8.09 &  7.81 &  0.08 &  29.8 &  23.2 &   2.3 \\
      121139 &  09 15 32.0   & +59 49 48.7 &   16.3 &   16.67 & $-$15.27 &   0.75 &   22.1 & 0.7 (8.3) & 0.89 &  8.29 & . & . & . & . & . \\
      132909 &  09 18 58.6   & +58 14 07.8 &   14.0 &   16.86 & $-$14.73 &   0.27 &   22.0 & 0.6 (8.3) & 0.40 &  7.61 &  8.26 &  0.09 &  63.3 &  59.2 &   1.4 \\
      136373 &  09 27 53.7   & +60 24 20.3 &   15.8 &   16.66 & $-$15.20 &   0.35 &   21.3 & 0.5 (6.5) & 0.49 &  7.84 &  7.53 &  0.09 &  26.9 &  21.3 &   3.3 \\
      276603 &  09 54 35.7   & +04 23 07.9 &   19.7 &   17.20 & $-$15.21 &   0.20 &   21.7 & 0.7 (6.8) & 0.57 &  7.72 &  8.25 &  0.07 &  50.8 &  51.2 &   6.6 \\
      232890 &  09 54 55.0   & +56 36 28.4 &   19.3 &   16.93 & $-$15.32 &   0.27 &   21.2 & 0.5 (5.3) & 0.36 &  7.78 &  8.20 &  0.07 &  61.7 &  56.5 &   1.9 \\
      169071 &  10 04 25.1   & +02 33 31.4 &   13.5 &   17.22 & $-$14.28 &   0.26 &   22.2 & 0.6 (8.7) & 0.48 &  7.43 &  7.05 &  0.20 &  12.9 &   5.5 &   3.4 \\
      278622 &  10 07 04.5   & +05 00 24.6 &   19.2 &   16.63 & $-$15.63 &   0.35 &   20.4 & 1.2 (12.9) & 0.56 &  8.07 &  8.13 &  0.08 &  32.8 &  29.5 &   6.0 \\
      262647 &  10 17 02.3   & +03 38 45.6 &   12.8 &   16.22 & $-$15.15 &   0.35 &   19.9 & 0.6 (9.5) & 0.55 &  7.86 &  7.56 &  0.14 &  31.1 &  27.2 &   7.4 \\
      133217 &  10 20 24.3   & +63 26 23.7 &   20.8 &   16.96 & $-$15.43 &   0.43 &   21.3 & 1.1 (10.8) & 0.17 &  8.04 &  8.25 &  0.06 &  63.3 &  55.3 &   2.2 \\
      565755 &  10 27 01.8   & +56 16 14.4 &   11.2 &   15.42 & $-$15.60 &   0.28 &   20.1 & 0.4 (7.8) & 0.56 &  7.96 &  8.25 &  0.11 &  47.2 &  46.5 &   2.7 \\
      114699 &  10 31 26.8   & +64 15 25.8 &   20.9 &   16.94 & $-$15.48 &   0.47 &   19.7 & 0.4 (4.2) & 0.62 &  8.15 & . & . & . & . & . \\
      274726 &  10 32 01.3   & +04 20 45.9 &   13.8 &   17.75 & $-$13.85 &   0.32 &   22.2 & 0.3 (4.8) & 0.56 &  7.36 &  7.77 &  0.09 &  42.8 &  41.4 &   6.3 \\
      317892 &  10 39 51.5   & +56 43 59.2 &   14.4 &   16.45 & $-$15.14 &   0.20 &   21.5 & 1.1 (15.6) & 0.35 &  7.68 &  8.80 &  0.09 &  70.0 &  65.0 &   1.5 \\
      119086 &  10 50 34.8   & +66 02 42.4 &   14.2 &   16.46 & $-$15.11 &   0.39 &   21.7 & 0.7 (10.4) & 0.35 &  7.91 &  7.83 &  0.10 &  37.4 &  30.8 &   2.9 \\
      237224 &  11 05 53.7   & +60 22 29.2 &   16.9 &   16.29 & $-$15.64 &   0.12 &   21.9 & 0.8 (9.7) & 0.42 &  7.75 &  8.84 &  0.08 &  71.5 &  68.7 &   3.0 \\
      280272 &  11 08 22.0   & +05 53 25.7 &   25.7 &   17.68 & $-$15.25 &   0.19 &   21.2 & 0.9 (6.9) & 0.19 &  7.73 &  8.48 &  0.06 &  62.9 &  54.9 &   3.8 \\
      246454 &  11 14 05.2   & +02 11 54.8 &   15.6 &   16.53 & $-$15.32 &   0.31 &   21.8 & 1.0 (12.8) & 0.28 &  7.83 &  8.14 &  0.08 &  66.3 &  59.7 &   7.4 \\
      319836 &  11 22 35.7   & +58 58 40.9 &   16.0 &   16.37 & $-$15.46 &   0.23 &   19.7 & 0.5 (6.5) & 0.45 &  7.89 &  8.19 &  0.09 &  45.4 &  41.0 &   1.8 \\
      327197 &  11 25 05.4   & +04 07 15.6 &   17.6 &   17.04 & $-$15.12 &   0.21 &   21.1 & 0.6 (7.1) & 0.31 &  7.68 &  8.58 &  0.06 &  74.2 &  68.4 &   1.7 \\
      327205 &  11 26 08.3   & +04 03 44.5 &   16.9 &   16.45 & $-$15.58 &   0.18 &   21.1 & 0.6 (7.0) & 0.62 &  7.88 &  8.08 &  0.06 &  39.0 &  39.0 &   3.1 \\
      323534 &  11 30 14.4   & +59 56 26.9 &   13.1 &   16.45 & $-$14.95 &   0.34 &   21.7 & 1.0 (15.4) & 0.58 &  7.63 &  8.27 &  0.10 &  55.4 &  57.2 &   1.6 \\
      330228 &  11 35 18.4   & +04 57 17.4 &   15.9 &   16.99 & $-$14.87 &   0.36 &   21.7 & 0.4 (5.4) & 0.59 &  7.75 &  7.95 &  0.08 &  35.4 &  33.4 &   6.0 \\
       72997 &  11 53 28.6   & $-$03 13 47.2 &   15.9 &   16.17 & $-$15.66 &   0.32 &   20.6 & 0.4 (5.8) & 0.62 &  8.00 &  8.13 &  0.08 &  49.8 &  52.0 &   2.6 \\
      325105 &  11 53 49.0   & +60 52 09.8 &   15.8 &   16.13 & $-$15.74 &   0.55 &   19.0 & 0.6 (8.0) & 0.61 &  8.30 & . & . & . & . & . \\
       64798 &  11 55 03.6   & $-$03 30 12.2 &   15.5 &   17.53 & $-$14.26 &   0.33 &   21.1 & 0.3 (3.6) & 0.71 &  7.47 &  7.33 &  0.08 &  28.5 &  28.7 &   4.1 \\
       67068 &  11 57 12.4   & $-$02 41 11.3 &   15.3 &   16.61 & $-$15.16 &   0.16 &   21.5 & 0.7 (9.3) & 0.32 &  7.61 &  8.01 &  0.08 &  25.4 &  18.1 &   2.4 \\
      328956 &  11 59 00.8   & +04 40 10.7 &   17.5 &   16.37 & $-$15.69 &   0.32 &   19.6 & 0.3 (4.0) & 0.67 &  8.03 &  7.87 &  0.07 &  35.2 &  35.9 &   4.3 \\
      535294 &  11 59 43.3   & +53 36 39.1 &   14.4 &   16.57 & $-$15.05 &   0.37 &   21.8 & 0.7 (9.4) & 0.56 &  7.83 &  7.52 &  0.10 &  22.7 &  17.5 &   4.2 \\
      238933 &  12 02 43.3   & +62 29 52.4 &   13.3 &   15.82 & $-$15.62 &   0.26 &   21.8 & 1.0 (14.8) & 0.26 &  7.96 &  8.34 &  0.10 &  61.2 &  54.1 &   1.9 \\
      166775 &  12 08 20.0   & +02 30 19.9 &   20.7 &   16.66 & $-$15.76 &   0.31 &   19.8 & 0.3 (3.0) & 1.00 &  8.05 &  8.32 &  0.06 &  62.7 &   . &   . \\
      158190 &  12 18 47.1   & +01 22 53.6 &   17.5 &   17.69 & $-$14.37 &   0.24 &   22.1 & 0.6 (7.5) & 0.34 &  7.39 &  7.80 &  0.06 &  27.4 &  20.2 &   3.6 \\
      296747 &  12 19 49.9   & +64 05 24.7 &   17.5 &   16.89 & $-$15.17 &   0.55 &   18.8 & 0.6 (6.5) & 0.53 &  8.06 & . & . & . & . & . \\
      172859 &  12 20 54.7   & +03 24 11.7 &   22.5 &   17.10 & $-$15.49 &   0.22 &   21.4 & 0.6 (5.2) & 0.45 &  7.86 &  8.18 &  0.05 &  56.5 &  53.3 &   2.2 \\
      149339 &  12 23 30.0   & +02 00 29.1 &   19.2 &   16.72 & $-$15.54 &   0.31 &   21.3 & 1.4 (14.8) & 0.15 &  7.96 &  8.25 &  0.08 &  62.5 &  54.5 &  10.3 \\
      258802 &  12 24 35.4   & +64 42 34.9 &   18.1 &   16.77 & $-$15.35 &   0.51 &   20.6 & 0.7 (8.4) & 0.30 &  8.13 & . & . & . & . & . \\
      148003 &  12 32 46.4   & +01 34 08.9 &   16.7 &   16.98 & $-$14.97 &   0.21 &   22.2 & 0.7 (8.2) & 0.49 &  7.55 &  8.14 &  0.07 &  30.8 &  25.7 &   1.4 \\
      329114 &  12 32 58.8   & +04 34 45.0 &   13.9 &   15.92 & $-$15.62 &   0.41 &   22.8 & 1.3 (19.2) & 0.45 &  7.83 &  8.30 &  0.08 &  51.0 &  47.3 &   1.9 \\
       66304 &  12 36 43.7   & $-$03 01 14.4 &   25.3 &   16.85 & $-$16.04 &   0.20 &   20.9 & 0.6 (4.9) & 0.44 &  8.09 &  8.23 &  0.08 &  56.7 &  53.2 &  18.7 \\
       67359 &  12 37 47.0   & $-$02 31 59.4 &   24.3 &   16.71 & $-$16.08 &   0.35 &   19.8 & 0.6 (4.7) & 0.47 &  8.23 & . & . & . & . & . \\
       73362 &  12 41 28.9   & $-$03 15 13.4 &   18.8 &   16.85 & $-$15.38 &   0.45 &   20.8 & 1.0 (11.2) & 0.15 &  7.98 &  8.04 &  0.06 &  62.8 &  54.8 &   2.5 \\
        6716 &  12 52 53.2   & $-$00 49 21.7 &   15.8 &   16.72 & $-$15.12 &   0.37 &   21.1 & 0.7 (8.7) & 0.32 &  7.86 &  7.53 &  0.08 &  31.2 &  24.1 &   4.1 \\
       32636 &  12 54 05.1   & $-$00 06 04.2 &   10.9 &   15.99 & $-$15.02 &   0.54 &   21.6 & 0.6 (11.3) & 0.93 &  7.98 &  7.10 &  0.13 &  28.2 &  53.5 &   9.0 \\
       45831 &  13 02 40.8   & +01 04 26.8 &   10.4 &   16.92 & $-$13.97 &   0.28 &   20.6 & 0.3 (5.9) & 0.70 &  7.30 &  7.20 &  0.15 &  35.0 &  37.4 &  16.5 \\
      148433 &  13 29 55.8   & +01 32 38.4 &   12.0 &   15.99 & $-$15.25 &   0.28 &   21.9 & 0.5 (9.1) & 0.72 &  7.80 &  7.94 &  0.10 &  33.7 &  36.2 &   3.4 \\
      346768 &  13 44 17.3   & +61 14 09.0 &   21.4 &   16.87 & $-$15.61 &   0.27 &   21.7 & 1.2 (11.7) & 0.31 &  7.88 &  7.90 &  0.06 &  38.8 &  31.8 &   3.8 \\
      370242 &  14 19 11.5   & $-$02 15 15.6 &   25.8 &   17.60 & $-$15.35 &   0.10 &   20.5 & 0.6 (5.1) & 0.34 &  7.63 &  8.19 &  0.04 &  43.3 &  36.9 &   3.7 \\
      381017 &  14 22 30.7   & $-$01 13 44.4 &   18.5 &   16.78 & $-$15.45 &   0.27 &   22.3 & 1.0 (11.2) & 0.31 &  7.94 &  8.00 &  0.06 &  44.9 &  38.2 &   2.5 \\
      371747 &  14 27 04.8   & $-$01 43 46.8 &   18.8 &   17.67 & $-$14.62 &   0.26 &   21.1 & 0.4 (3.9) & 0.46 &  7.54 &  7.82 &  0.06 &  35.1 &  30.0 &   3.6 \\
      178719 &  14 38 22.6   & +04 36 47.8 &   17.8 &   17.35 & $-$14.81 &   0.34 &   21.1 & 0.5 (5.6) & 0.35 &  7.73 &  7.86 &  0.08 &  72.0 &  67.1 &   7.4 \\
      301586 &  14 41 33.7   & +03 29 48.0 &   17.7 &   16.50 & $-$15.61 &   0.52 &   19.3 & 1.0 (11.8) & 0.49 &  8.20 &  7.40 &  0.08 &  34.6 &  30.0 &  14.7 \\
      307723 &  14 46 20.2   & +04 43 58.4 &   16.4 &   17.56 & $-$14.40 &   0.18 &   22.0 & 0.5 (5.8) & 0.35 &  7.33 &  7.80 &  0.06 &  34.1 &  27.4 &   4.6 \\
      377686 &  14 56 20.3   & $-$02 45 41.8 &   19.4 &   17.96 & $-$14.53 &   0.15 &   22.2 & 0.4 (4.6) & 0.71 &  7.44 &  7.68 &  0.06 &  27.7 &  27.5 &   3.9 \\
      373377 &  15 09 33.6   & $-$01 01 17.5 &   19.1 &   17.09 & $-$15.28 &   0.40 &   12.9 & 1.7 (18.0) & 0.53 &  7.97 & . & . & . & . & . \\
      284191 &  15 13 06.3   & +56 58 08.8 &   11.4 &   15.77 & $-$15.33 &   0.46 &   22.4 & 0.9 (15.9) & 0.43 &  8.11 &  8.41 &  0.10 &  53.1 &  49.1 &   1.6 \\
      379759 &  15 32 39.2   & $-$01 36 06.2 &   26.4 &   17.80 & $-$15.48 &   0.29 &   21.5 & 0.9 (6.8) & 0.78 &  7.93 &  8.89 &  0.04 & 107.7 & 156.1 &  12.6 \\
      377995 &  15 38 29.6   & $-$02 23 46.2 &   31.1 &   17.82 & $-$15.84 &   0.11 &   21.3 & 0.8 (5.1) & 0.80 &  7.89 &  8.73 &  0.04 &  48.6 &  67.0 &   2.2 \\
      340807 &  16 09 38.8   & +45 16 42.8 &   19.8 &   16.55 & $-$15.76 &   0.29 &   22.4 & 1.8 (18.3) & 0.49 &  7.99 &  8.77 &  0.05 &  67.5 &  67.1 &   2.2 \\
      208160 &  20 41 56.9   & $-$05 17 56.5 &   34.0 &   17.89 & $-$15.70 &   0.46 &   19.8 & 0.3 (2.1) & 0.84 &  8.14 &  8.12 &  0.05 &  51.2 &  78.8 &   8.9 \\
      398936 &  21 15 37.4   & $-$00 24 31.3 &   32.0 &   17.37 & $-$16.10 &   0.63 &   22.4 & 0.9 (5.5) & 0.93 &  8.39 &  8.07 &  0.07 &  21.1 &  34.4 &   7.0 \\
      217927 &  21 20 06.0   & +11 55 06.5 &   11.7 &   15.72 & $-$15.63 &   0.38 &   19.5 & 0.3 (5.6) & 0.55 &  8.11 &  8.31 &  0.07 &  79.2 &  83.7 &   1.3 \\
      209838 &  21 22 02.3   & +09 53 10.6 &   28.9 &   17.19 & $-$16.04 &   0.36 &   19.5 & 0.4 (2.6) & 0.79 &  8.22 &  7.82 &  0.13 &  43.4 &  56.4 &  34.2 \\
      189157 &  21 48 39.9   & $-$06 50 11.4 &   28.7 &   17.45 & $-$15.71 &   0.29 &   21.0 & 0.6 (4.3) & 0.55 &  8.03 & . & . & . & . & . \\
      181782 &  22 21 00.1   & $-$09 38 30.2 &   27.7 &   17.40 & $-$15.72 &   0.03 &   22.0 & 1.1 (8.1) & 0.22 &  7.80 &  8.71 &  0.04 &  68.6 &  61.1 &   2.4 \\
      421204 &  22 30 36.8   & $-$00 06 37.0 &   15.3 &   17.06 & $-$14.83 &   0.11 &   11.0 & 0.4 (5.1) & 0.85 &  7.37 &  7.84 &  0.08 &  39.1 &  57.4 &   7.4 \\
      186498 &  22 39 25.1   & $-$08 45 57.7 &   38.7 &   17.85 & $-$16.04 &   0.30 &   21.6 & 1.1 (5.9) & 0.62 &  8.19 &  8.30 &  0.05 &  53.9 &  57.6 &   7.6 \\
      392224 &  22 59 24.7   & $-$10 29 16.2 &   31.6 &   17.56 & $-$15.84 &   0.34 &   22.4 & 1.2 (7.6) & 0.70 &  8.14 &  8.34 &  0.07 &  30.6 &  31.1 &   5.0 \\
      410317 &  23 01 20.5   & $-$00 55 33.3 &   28.7 &   17.20 & $-$16.00 &   0.31 &   21.5 & 0.9 (6.1) & 0.34 &  8.16 &  8.35 &  0.09 &  64.4 &  58.8 &  21.2 \\
      213776 &  23 05 11.2   & +14 03 47.4 &   14.3 &   16.52 & $-$15.57 &   0.31 &   21.1 & 0.3 (4.7) & 0.72 &  7.95 &  7.90 &  0.07 &  52.1 &  62.0 &   4.9 \\
      183623 &  23 05 56.3   & $-$10 02 57.0 &   21.3 &   17.07 & $-$15.47 &   0.11 &   21.5 & 0.5 (5.2) & 0.64 &  7.67 &  8.33 &  0.07 &  40.1 &  41.0 &   5.0 \\
      215437 &  23 06 15.1   & +14 39 27.4 &   14.2 &   16.88 & $-$15.19 &   0.27 &   22.5 & 0.5 (6.8) & 0.87 &  7.77 &  8.48 &  0.07 &  54.7 &  92.0 &   1.8 \\
      633306 &  23 28 12.3   & $-$01 03 46.2 &   24.5 &   17.43 & $-$15.38 &   0.13 &   21.8 & 0.8 (6.5) & 0.53 &  7.67 &  8.35 &  0.06 &  41.9 &  39.3 &   4.6 \\
      421787 &  23 31 02.1   & $-$00 07 55.0 &   24.1 &   17.16 & $-$15.63 &   0.36 &   21.7 & 1.0 (8.5) & 0.29 &  8.02 &  8.60 &  0.05 &  61.6 &  54.9 &   2.3 \\
\enddata \tablecomments{Galaxy names are based on position in the
\citet{blanton04b} low luminosity SDSS catalog.  Absolute magnitudes
and $g-r$ colors have been corrected for Galactic extinction based on
\citet{schlegel98a}.  Distances are estimated based on a model of the
local velocity field \citep{willick97a} and assume H$_{0} = 70
h_{70}$\,Mpc \kms.  The effective surface brightness ($\mu_{{\rm
eff},r}$), effective half-light radius ($r_{\rm eff,r}$) and axis
ratio ($b/a$) are measured from the SDSS $r$-band images.  Stellar
masses are calculated based on \citet{bell03a} and a \citet{kroupa93a}
initial mass function; HI masses are based on our Arecibo and GBT
observations.  The HI velocity half-width, W20/2, is the observed
uncorrected 20\% peak HI half-width.  The quantity W20$_{i,t}$/2 is
the HI half-width corrected for inclination and turbulence as
described in \S\,\ref{ssec_tf}.  Errors on the HI mass
($\sigma_{M_{\rm HI}}$) and HI line-width ($\sigma_{W20_{i,t}/2}$)
were calculated from the scatter in the mean quantities recovered from Monte Carlo simulations.  Galaxies which were not detected in the radio
do not have HI properties listed in this table.}
\end{deluxetable}
\clearpage
\begin{deluxetable}{lrrc}
\tablewidth{0pt}
\tablecaption{Tully-Fisher Relation Fit Parameters}
\tablehead{
\colhead{Galaxy Sample} &
\colhead{$a$} & 
\colhead{$b$} & 
\colhead{$\sigma$}  }
\startdata
Stellar mass ~~~(current sample) & $-$0.14 $\pm$ 0.02 & 0.43 $\pm$ 0.06 & 0.12 \\
Baryonic mass (current sample) & $-$0.32 $\pm$ 0.05 & 0.53 $\pm$ 0.08 & 0.13 \\
$M_I$ ~~~~~~~~~~~~~~(full sample) & 0.37 $\pm$ 0.01 & $-$0.09 $\pm$ 0.01 & 0.11 \\
Baryonic mass (full sample) & $-$0.19 $\pm$ 0.02 & 0.27 $\pm$ 0.01 & 0.10 \\
\enddata 
\tablecomments{ Tully-Fisher linear regression fit parameters for
Equation \ref{tffit_eqn}. The zero-point $a$ is in units
$\log_{10}($W20$_{i,t}/2/(50$ km s$^{-1}))$.  The slope $b$ is in
units $\log_{10}($mass$/10^8 h_{70}^{-2} M_\odot)$ per
$\log_{10}($W20$_{i,t}/2/(50$ km s$^{-1}))$. In the $I$-band
Tully-Fisher relationship we replace $\log_{10}($mass$/10^8
h_{70}^{-2} M_\odot)$ with $M_I -5 \log_{10} h_{70} + 20$. The
quantity $\sigma$ represents the logarithmic (base 10) scatter around
each fit.  The first two rows ('current sample') are fits to the
galaxy data presented in this paper; the last two rows ('full sample')
are fits to this sample plus data from the literature as desribed in
\S\,\ref{ssec_tf}.\label{tffit_table}}
\end{deluxetable}

\clearpage

\end{document}